\pdfoutput=1
\documentclass[journal,10pt]{IEEEtran}
\usepackage{amsmath,amssymb,amsfonts,amsthm,booktabs,array,tabularx,graphicx,cite,url,xurl}
\usepackage[T1]{fontenc}
\usepackage{accsupp}
\usepackage{xcolor}
\usepackage{algorithm}
\usepackage{needspace}
\usepackage{multirow}
\usepackage{algpseudocode}
\newcommand{\algblockhead}[2]{\Statex\hspace*{-\leftmargin}\rlap{\raisebox{\dimexpr\ht\strutbox+1.5pt\relax}{\rule{\dimexpr\linewidth+\leftmargin\relax}{0.3pt}}}\rule{0pt}{\dimexpr\ht\strutbox+4pt\relax}\textbf{#1}\hfill#2}
\usepackage[hidelinks]{hyperref}
\newcommand{\figFiveBiasSE}{0.026}
\newcommand{\figFiveRmseSE}{0.026}
\newcommand{\figSixRmseSE}{0.22}
\newcommand{\figSensorBiasSE}{0.040}%
\newcommand{\figSensorRmseSE}{0.039}%
\newcommand{\figSensorWassSE}{0.023}%
\hypersetup{pdftitle={Exponential Quantum Memory Advantage for Late-Graph Smoothness and Spectral Estimation},pdfauthor={Burhan Gulbahar}}
\newtheorem{theorem}{Theorem}\newtheorem{lemma}{Lemma}\newtheorem{proposition}{Proposition}
\newtheorem{corollary}{Corollary}
\theoremstyle{definition}\newtheorem{definition}{Definition}\newtheorem{remark}{Remark}
\makeatletter\newcommand{\sectionlevellabel}[1]{\begingroup\protected@edef\@currentlabel{\thesection}\label{#1}\endgroup}\makeatother
\newcommand{\ket}[1]{\lvert#1\rangle}\newcommand{\bra}[1]{\langle#1\rvert}
\newcommand{\cE}{\mathcal E}\newcommand{\E}{\mathbb E}\newcommand{\bz}{\mathbf z}\newcommand{\bL}{\mathbf L}
\newcommand{\bH}{\mathbf H}\newcommand{\tr}{\operatorname{tr}}\newcommand{\wmax}{w_{\max}}

\begin{document}
\bstctlcite{BSTcontrol}
\title{Exponential Quantum Memory Advantage for \\ Late-Graph Smoothness and Spectral Estimation}
\author{Burhan~G\"ulbahar\thanks{B. G\"ulbahar is with the Department of Electrical and Electronics Engineering, \protect\BeginAccSupp{method=hex,unicode,ActualText=00590061015F00610072}Ya\c{s}ar\protect\EndAccSupp{} University, \protect\BeginAccSupp{method=hex,unicode,ActualText=0130007A006D00690072002C}\.{I}zmir,\protect\EndAccSupp{} T\"urkiye (e-mail: burhan.gulbahar@yasar.edu.tr).}}
\markboth{}{G\"ulbahar: Exponential Quantum Memory Advantage for Late-Graph Smoothness and Spectral Estimation}
\maketitle

\begin{abstract}
Graph signal processing tests how smoothly the decisions of next-generation massive sensor networks with $N$ nodes vary over their neighbour graph. Memory becomes the bottleneck when the neighbour graph arrives only after data erasure, a late-graph setting motivated by dispersed sensor networks and telemetry. We analyse three collection scenarios, reliable rounds, continuous broadcast and ordered capture, and test the Dirichlet energy $\mathcal E_G=\mathbf z^{\mathsf T}\mathbf L\mathbf z/N$ of the fused decision map $\mathbf z\in\{\pm1\}^N$, with late-graph Laplacian $\mathbf L$, against two thresholds. Hidden Matching uses $O(\log N)$ qubits on fused bits, but in reliable rounds and continuous broadcast, where nodes report $b$ sub-bits separately, a classical record buffer before it uses $O(N b)$ bits; classical methods need $\Omega(\sqrt N)$ bits in the worst case. The recently introduced Quantum Oracle Sketching builds quantum oracles from random data streams without storing them. We combine coherent loading of separately reported sub-bits, exact in reliable rounds and by Quantum Oracle Sketching in continuous broadcast, with nonlinear fusion and Hidden Matching readout. The register count $R$ does not grow with $N$, and the memory is $O(\log N)$ logical qubits and classical bits, at fixed $b$, graph bounds, thresholds and error targets: an exponential memory advantage over every classical receiver and over Hidden Matching with that buffer. We prove unbiasedness in reliable rounds, preparation bounds in continuous broadcast, classical lower bounds and, for a planner-supplied multi-hop coefficient stream, spectral error bounds. Broadcast simulations match analytic averages at $N=1024$, without resolved bias growth up to $N=2^{20}$. 
Fault-tolerant hardware, gate rates and, in continuous broadcast, report counts remain open.
\end{abstract}

\section{Introduction}\label{sec:intro}
\IEEEPARstart{G}{raph} signal processing studies signals on the vertices of a graph whose edges join neighbours \cite{Shuman2013,Ortega2018,Sandryhaila2013}. Take a signal $\bz\triangleq(z_0,\ldots,z_{N-1})^{\mathsf T}$ on an undirected graph $G\triangleq([N],E,w)$ with vertices $[N]\triangleq\{0,\ldots,N-1\}$, edges $E$ and weights $w_e>0$. The graph quadratic form measures smoothness \cite{Shuman2013}. Dividing it by $N$ gives the Dirichlet energy:
\begin{equation}
\cE_G(\bz)\triangleq\frac{\bz^{\mathsf T}\bL\bz}{N}=\frac1N\sum_{e=\{u,v\}\in E}w_e\,(z_u-z_v)^2,\label{eq:energyintro}
\end{equation}
where $\bL\triangleq\mathbf D-\mathbf W$ is the graph Laplacian with adjacency matrix $\mathbf W$ and degree matrix $\mathbf D$ \cite{Shuman2013}. Small values mean a smooth signal, large values a rough one; under such a smoothness model, high-pass filtering and thresholding detect sensor-network outliers \cite{Ortega2018}. Each site $x$ fuses its $b$ one-bit readings $\mathbf s_x$ into a status sign $z_x\in\{\pm1\}$ \cite{ChairVarshney1986}. Each term of \eqref{eq:energyintro} is therefore $4w_e$ for disagreeing neighbours and $0$ otherwise. Rough maps motivate anomaly screening.

Next-generation sensor and public networks may have a massive number $N$ of nodes, and report memory grows with $N$. In reliable rounds (RR), an aircraft drops pods of $b$ battery-free one-bit devices, such as ambient-backscatter tags \cite{Liu2013} or wind-dispersed environmental sensors \cite{Iyer2022}, over a disaster area; we assume a base station or satellite receives every bit exactly once in each of two rounds, a contract motivated by the acknowledged uplinks of long-range low-power networks \cite{LoRaWAN2020}, after which the devices erase their bits. In continuous broadcast (CB), sources such as Bluetooth beacons on moving assets \cite{Jeon2018} or Internet traffic telemetry sampled at equal rates per cell \cite{Duffield2004} transmit, the receiver keeps nothing it hears, and reports arrive uniformly at random over (site, bit) cells. The neighbour graph in both cases comes from aerial images of the dispersed devices \cite{Kim2021} or flow logs of network monitoring \cite{Hofstede2014}. It is known only after erasure. We call it the late graph and call erasure the cutoff. Late graphs arise when the geometry or the coupling of the sites is measured by a separate, slower process than the readings, such as a survey after dispersal, aerial disaster imaging or the offline assembly of traffic logs, while memory or retention limits erase raw reports before that process ends. Denser deployments with shorter data retention can make such cutoffs stricter. No position information reaches the receiver before this cutoff. An early graph, known before the reports, would let the receiver keep only the bits at the endpoints of a few sampled edges. Any two sites may prove to be neighbours on a late graph after their reports are gone.

Zhao et al.\ \cite{ZhaoQOS} introduced quantum oracle sketching (QOS), which builds quantum query oracles from randomly streamed classical data. Each sample applies a small phase gate to a register and is discarded. The register undergoes the data's phase oracle on average without stored samples or quantum random-access memory (QRAM) \cite{QRAM2008}. QOS is proposed for sparse linear systems, binary classification and dimension reduction, where a polylogarithmic-size quantum machine matches exponentially larger classical ones.

\begin{figure}[!t]
\centering
\includegraphics{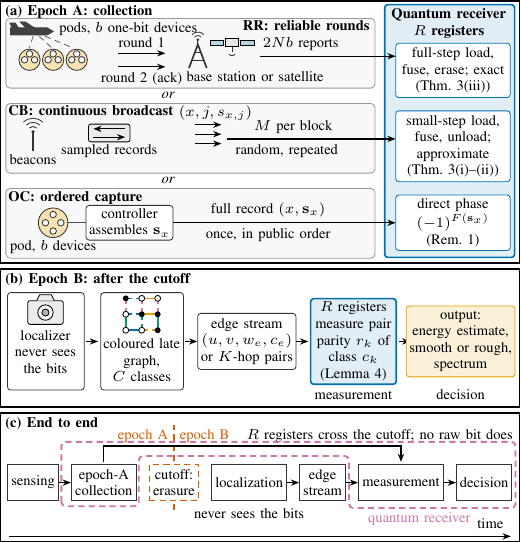}
\caption{Late-graph estimation from erased one-bit sensor reports. (a) Epoch~A: RR, CB or OC feeds one quantum receiver (load, fuse and erase or unload; direct phase in OC). (b) Epoch~B: a localizer that never sees the bits streams the coloured late graph once; dashed edges join disagreeing endpoints. For the spectrum, the planner instead streams the vertex pairs within $K$ hops with their coefficients (Theorem~\ref{thm:spectrum}). The measurement block uses the $R$ registers to measure pair parities. The decision block outputs an energy estimate, a smooth/rough decision or the spectrum. (c) End to end: only the registers cross the cutoff; the dashed outline marks the quantum receiver.}\label{fig:overview}
\end{figure}

We combine QOS with Hidden Matching (HM) \cite{BJK2004,GKKRW,KPV2024,KPV2025}, a one-way communication problem in which a $\log_2 N$-qubit state reveals one pair parity of a later-chosen matching, whereas classical messages need $\Omega(\sqrt N)$ bits. A public fusion rule $F$ maps $b$ bits of a pod, called sub-bits, to one status; each pod is a site $x$ of $N$ sites with record $\mathbf s_x$. We test whether the Dirichlet energy of the fused map on the late graph lies below one threshold or above a higher one. Our receiver keeps $R$ registers. It loads the sub-bit reports coherently: exactly in RR and by QOS in CB. Then, it fuses each site's sub-bits nonlinearly in the label qubits by load, fuse and unload or erase steps. Finally, it reads pair parities on the colour classes of the late graph as in HM. Figs.~\ref{fig:overview}(a)--(c) show the collection of reports in epoch A, the pair-parity readout of the coloured late graph in epoch B and the end-to-end pipeline; only the registers cross the cutoff. The receiver retains $O(\log N)$ logical qubits and $O(\log N)$ bits of classical control for fixed record width, class count, weight bound, thresholds and error targets. HM pair sketches with a classical record buffer use $O(Nb)$ bits in RR and CB. Every classical receiver needs $\Omega(\sqrt N)$ bits in the worst case, giving an exponential memory separation. Our comparison case, ordered capture (OC), supplies each whole record once. HM alone \cite{GKKRW} then uses $O(\log N)$ logical qubits and $O(\log N)$ classical bits, with one record buffered at a time (Remark~\ref{rem:direct}). An early graph lets classical and quantum receivers reach logarithmic memory in all three scenarios. Sampling $O(1)$ edges and keeping their endpoint records takes $O(\log N)$ classical bits at fixed accuracy (Section~\ref{sec:classical}). The separation therefore rests on a late, unpredictable graph. Table~\ref{tab:scenarios} lists the three scenarios. The proposed exponential advantage is in memory, an information resource, and is not a claim of near-term hardware gain; RR takes twice the classical reports and CB many more. Table~\ref{tab:compare} compares the peak classical and quantum memory and the reports per raw bit. To our knowledge, this is the first study of graph-signal smoothness and spectral estimation on a late graph from separately reported, erased sensor bits with logarithmic quantum and classical memory.

\begin{table}[!t]
\caption{Three epoch-A collection scenarios}\label{tab:scenarios}
\centering\scriptsize\setlength{\tabcolsep}{1.5pt}\renewcommand{\arraystretch}{1.0}\setlength{\extrarowheight}{1pt}
\begin{tabularx}{\columnwidth}{|l|>{\raggedright\arraybackslash}X|>{\raggedright\arraybackslash}X|l|}\hline
	Scenario & Collection & Receiver & Result\\\hline
	RR, reliable rounds & two acknowledged rounds; erase & full-step load, fuse, erase; exact & Thm.~\ref{thm:prep}(iii)\\\hline
	CB, continuous broadcast & random, repeated reports & small-step load, fuse, unload; approximate & Thm.~\ref{thm:prep}(i)--(ii)\\\hline
	OC, ordered capture & full record once & direct phase & Rem.~\ref{rem:direct}\\\hline
\end{tabularx}
\par\smallskip\parbox{\columnwidth}{\scriptsize Conditions: Section~\ref{sec:contracts}; examples: Section~\ref{sec:intro}; memory and reports: Table~\ref{tab:compare}.}
\end{table}

\begin{table}[!t]
\caption{Peak memory and reports per bit in RR and CB}\label{tab:compare}
\centering\scriptsize\setlength{\tabcolsep}{1pt}\renewcommand{\arraystretch}{1.0}\setlength{\extrarowheight}{2pt}
\begin{tabularx}{\columnwidth}{|l|>{\raggedright\arraybackslash}X|c|c|c|c|}\hline
\multicolumn{2}{|l|}{} & & & \multicolumn{2}{c|}{Reports per raw bit}\\\cline{5-6}
\multicolumn{2}{|l|}{Method} & Classical (bits) & Quantum (qubits) & RR & CB\\\hline
\multicolumn{2}{|l|}{Store all} & $O(Nb)$ & -- & $O(1)$ & $O(\log(Nb/\delta))$\\\hline
\multicolumn{2}{|l|}{Site cache} & $O(\sqrt N\log N)$ & -- & $O(1)$ & $O(\log(s_{\rm c}b/\delta_{\rm miss}))$\\\hline
\multicolumn{2}{|l|}{Any classical} & $\Omega(\sqrt N)$ & -- & -- & --\\\hline
\multicolumn{2}{|l|}{\parbox[t]{0.2\columnwidth}{\raggedright Buffer + HM \cite{GKKRW,KPV2024,KPV2025}\strut}} & $O(Nb)$ & $O(R\log N)$ & $O(1)$ & $O(\log(Nb/\delta))$\\\hline
\multicolumn{2}{|l|}{QRAM \cite{QRAM2008}} & \multicolumn{2}{c|}{$\Theta(Nb)$} & $O(1)$ & $O(\log(Nb/\delta))$\\\hline
\multirow{2}{*}{\textbf{This work}} & shared & $O(R+\log N)$ & $O(R(\log N+b))$ & $O(1)$ & $O(Rb/\delta_p)$\\\cline{2-6}
 & serial & $O(R+\log N)$ & $O(R\log N+b)$ & -- & $O(R^2b/\delta_p)$\\\hline
\end{tabularx}
\par\smallskip\parbox{\columnwidth}{\scriptsize $R$: registers; $b$: record width; $s_{\rm c}$: cached sites; $\delta$, $\delta_{\rm miss}$: coverage failure and cache miss for IID draws; $\delta_p$: preparation distance; sufficient budgets; classical bits at fixed $b$, $C$, $\wmax$, thresholds and error targets. Test of Theorem~\ref{thm:dirichlet}(iii) alone, serial CB: $O(Rb\lambda_N/\gamma)$ (Proposition~\ref{prop:serial}).}
\end{table}

\subsection{Related Work}\label{sec:related}
Graph signal processing provides smoothness, filtering and polynomial-kernel methods \cite{Shuman2013,Ortega2018,Sandryhaila2013,Hammond2011}. Spectral estimation \cite{Marques2017,Perraudin2017} and sampling \cite{Tanaka2020} use the graph during processing. Matrix sketches \cite{FD2016} do not cover nonlinear fusion with late-chosen pairs. HM supplies the pair readout \cite{BJK2004}. Its partial-matching form $\alpha$PM gives the bounded-error classical obstruction and a one-pass streaming protocol \cite{GKKRW}. Pair sketches extend this protocol to counting labelled matching edges \cite{KPV2024} and to classical algorithms with a quantum subroutine \cite{KPV2025}. The late graph has a counterpart in one-way quantum communication: Gilboa et al.\ \cite{Gilboa2024} use a generalization of $\alpha$PM \cite{Doriguello2020} to prove an exponential advantage for quadratic graph networks in which Alice holds the node features and Bob the graph. Their cost is the message size: Alice is given all features as her input and local memory is not counted, so a short message alone does not imply a small memory. Our cost is the peak receiver memory: each report is applied as it arrives and discarded, separately reported sub-bits are fused, and the loading itself stays within logarithmic memory for fixed record width, register count and error targets (Theorem~\ref{thm:prep}(iv)). Recent work builds HM states by QOS phase updates from randomly ordered reports interleaved with matching requests \cite{Voronova2026}; our reports precede the late graph. We extend such loading to nonlinear fusion of isolated sub-bits. Each load or unload is one inner-product phase call. Indexed-bit subsampling used through $2b$ controlled single-bit calls \cite{ZhaoQOS} has a sufficient budget larger than those of Theorem~\ref{thm:prep} by a factor of order $b$. QOS estimates quadratic forms of a fixed matrix with a polylogarithmic measurement circuit. It also outputs polylogarithmic classical models for classification and dimension reduction \cite{ZhaoQOS} and separates memory for linear-system sampling \cite[App.~E]{Chakrabarti2026}. Here the matrix is the Laplacian of a graph streamed once after the cutoff. A classical summary for the test needs $\Omega(\sqrt N)$ bits in the worst case, so the registers themselves cross the cutoff. HM-based streaming separations can rely on adversarial order \cite{ZhaoQOS}, although quantum space advantages can persist in random data streams \cite{BDJ2026}. Our separation holds for CB reports in random order because the graph arrives after the cutoff. Trapped ions have run streaming HM \cite{Niroula2025} and shown an unconditional memory separation for another one-way communication task \cite{Kretschmer2025}.

\needspace{4\baselineskip}%
\subsection{Contributions}
This work builds on three known tools: QOS loading \cite{ZhaoQOS}, the HM pair readout with its one-way bound \cite{BJK2004,GKKRW} and the kernel polynomial method (Section~\ref{sec:kpm}). It contributes:
\begin{itemize}\setlength{\itemsep}{1pt}
	\item \emph{Late-graph memory model and coherent loading:} a cutoff between two epochs; three collection scenarios; a cost model comparing bits with qubits (Section~\ref{sec:model}) and gate counts (Section~\ref{sec:complexity}); exact loading in RR, QOS loading in CB (Theorem~\ref{thm:prep}) and a direct phase in OC (Remark~\ref{rem:direct}).
	\item \emph{Smoothness and spectral estimators with logarithmic memory:} smooth/rough tests; estimates of the Dirichlet energy and kernel functionals (Theorems~\ref{thm:dirichlet}--\ref{thm:kernel}); spectral distributions from planner-supplied $K$-hop Chebyshev coefficients (Theorem~\ref{thm:spectrum}).
	\item \emph{Exponential memory separation and its source:} every classical receiver needs $\Omega(\sqrt N)$ boundary bits in the worst case, also for the spectral task and even when only the cross edges of a public backbone are late (Theorems~\ref{thm:classical} and~\ref{thm:ring}, Corollaries~\ref{cor:separation} and~\ref{cor:specclassical}). For the smoothness test, a site cache of $O(\sqrt N\log N)$ bits nearly attains this bound, and an early graph needs only $O(\log N)$ bits (Section~\ref{sec:classical}).
	\item \emph{QOS extended to nonlinear fusion and late-graph readout:} QOS estimates quadratic forms of a fixed matrix \cite{ZhaoQOS}, and HM reads fused bits \cite{GKKRW}. Here separately reported sub-bits are loaded and unloaded by one inner-product phase call each and fused nonlinearly in the label qubits, with sufficient budgets for $R$ registers on shared or separate reports (Theorem~\ref{thm:prep}); HM then reads the fused map on a graph that arrives after the cutoff (Theorem~\ref{thm:dirichlet}).
	\item \emph{Simulations:} an $R=112$ certificate gives false-alarm and miss probabilities below $0.027$ and $0.043$ for ideal registers on the hard family at every $N\ge128$ (Appendix~\ref{app:cert}). End-to-end CB runs match the analytic tape averages (within $1.8$ standard errors on the sensor graph). They show no resolved bias growth up to $N=2^{20}$, consistent with the $N$-independent bound of Theorem~\ref{thm:prep} (Section~\ref{sec:numerics}). Gate-level circuits reach $N=2^{19}$. Their decision errors fall below $0.05$ by $R=256$ at every size up to $2^{16}$, against the sufficient $R=819$.
\end{itemize}

\noindent\textit{Notation:}\sectionlevellabel{sec:notation} Bold lowercase, uppercase, Greek: vectors, matrices, density matrices; $\log\triangleq\log_2$; $\|{\cdot}\|$, $\|{\cdot}\|_1$, $\|{\cdot}\|_F$: Euclidean\slash spectral, $\ell_1$\slash trace, Frobenius; $D_{\rm tr}(\boldsymbol\rho,\boldsymbol\sigma)\triangleq\frac12\|\boldsymbol\rho-\boldsymbol\sigma\|_1$; $\operatorname{Tr}_L$: label partial trace; $\mathbf 1\{\cdot\}$, $\mathbf1[\cdot]$: indicators; $\E[\cdot]$: expectation, with a subscript naming the random variable when needed ($\E_\omega[\cdot]$: average over the tape $\omega$ of Section~\ref{sec:loading}); $\oplus$: sum modulo two; $\ket x$: basis state of address $x$; $c_e$: colour class of edge $e$; $c_k$, $r_k$: class and parity of register $k$; $K$: polynomial order.

Section~\ref{sec:qos} recalls QOS; Sections~\ref{sec:model}--\ref{sec:spectrum} give the model, the receiver, the classical bounds and the spectral estimators, and Section~\ref{sec:numerics} the simulations. The proofs are in Appendices~\ref{app:prep} (Theorem~\ref{thm:prep}, Proposition~\ref{prop:serial}), \ref{app:est} (Lemma~\ref{lem:outcome}, Theorems~\ref{thm:dirichlet} and~\ref{thm:kernel}), \ref{app:classical} (Theorems~\ref{thm:classical} and~\ref{thm:ring}), \ref{app:spectrum} (Theorem~\ref{thm:spectrum}, Corollary~\ref{cor:specclassical}) and~\ref{app:lemmas} (Lemmas~\ref{lem:dirichlet} and~\ref{lem:colour}).

\section{Quantum Oracle Sketching Background}\label{sec:qos}

QOS builds a quantum query to a classical table from streamed samples without stored data \cite{ZhaoQOS}. Its data-access model lets a machine read samples one by one, update its memory after each, and estimate a property of their generating process. 
Numerical evidence on real data sets in \cite{ZhaoQOS} concerns QOS machines with fewer than $60$ logical qubits. It suggests six orders of magnitude less memory than sparse-matrix and QRAM-based methods and four orders less than tested classical streaming methods at matched performance.

Consider a Boolean table $f$ with $D$ cells and a register of $\lceil\log D\rceil$ qubits whose basis states $\ket c$ label the cells. Phase oracle $\mathbf O_f\triangleq\sum_c(-1)^{f(c)}\ket c\!\bra c$ (identity on unused basis states) maps the uniform superposition $D^{-1/2}\sum_c\ket c$ to $D^{-1/2}\sum_c(-1)^{f(c)}\ket c$; the register may start in any state. For $M$ independent and identically distributed (IID) uniform samples $(c_t,f(c_t))$, QOS applies $\mathbf V_t\triangleq\exp(i\vartheta f(c_t)\ket{c_t}\!\bra{c_t})$ with $\vartheta\triangleq\pi D/M$ and then discards the sample \cite{ZhaoQOS}. The gates commute, so $\ket c$ ends with phase $\vartheta n_cf(c)$, where $n_c$ counts the samples of $c$. Since $\E[n_c]=M/D$, the mean phase is $\pi f(c)$, the phase of $\mathbf O_f$. The residual phase $\vartheta(n_c-M/D)$ of an active cell has root-mean-square value at most $\pi\sqrt{D/M}$ because $\operatorname{Var}[n_c]\le M/D$. The sampled cells $c_1,\ldots,c_M$ form the tape; since it is discarded, the register undergoes the tape-averaged channel $\boldsymbol\rho\mapsto\E[\mathbf V_M\cdots\mathbf V_1\boldsymbol\rho\mathbf V_1^\dagger\cdots\mathbf V_M^\dagger]$, with expectation over the tape \cite{ZhaoQOS}. Its diamond-norm distance from the channel of $\mathbf O_f$ is a worst case over all inputs, including those entangled with other qubits. This distance is at most $\epsilon_\diamond$ once $M\ge2\pi^2D/\epsilon_\diamond$. Measurement statistics then change by at most $\epsilon_\diamond/2$ \cite[Thm.~D.12, Eq.~(D55)]{ZhaoQOS}. Hence $O(DQ^2/\epsilon_\diamond)$ samples suffice for $Q$ queries at total error $\epsilon_\diamond$; the sketch uses only the register and stores no sample, whereas QRAM stores the whole table \cite{ZhaoQOS}. Section~\ref{sec:loading} applies these steps to our table of $D=Nb$ sub-bits. Each register has a $d$-qubit address with $N=2^d$ and a $b$-qubit label. It starts in the state $\ket{+}^{\otimes d}\ket{0^b}$ and targets the phase state $\ket{\psi_{\bz}}$ of the fused map.

\section{Late-Graph Sensing and Sub-Bit Loading}\label{sec:model}
\begin{figure}[!t]
\centering
\includegraphics{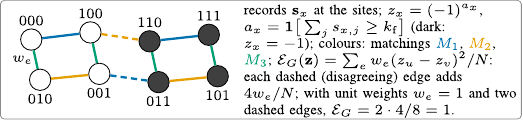}
\caption{Example for $N=8$, $b=3$, fusion threshold $k_{\rm f}=2$ and unit weights $w_e=1$: records $\mathbf s_x$ and fused signs $z_x$ (white $+1$, dark $-1$) on the late graph $G$, with edges coloured by matching and dashed where the signs disagree.}\label{fig:model}
\end{figure}

\subsection{Fused decision maps and late graphs}\label{sec:fused} A public rule fuses each site's sub-bits into one sign, and the Dirichlet energy of these signs on the late graph measures how often neighbours disagree. For threshold fusion, $k_{\rm f}\in\{1,\ldots,b\}$ is the required number of positive sub-bits. Fig.~\ref{fig:model} illustrates the objects of this subsection on an example.
There are $N=2^d$ sites with static records $\mathbf s_x\in\{0,1\}^b$ of sub-bits $s_{x,j}$. A public fusion rule gives the status $a_x\triangleq F(\mathbf s_x)$ and the fused sign $z_x\triangleq(-1)^{a_x}$; the running example is threshold fusion, translated from the counting rule in \cite{ChairVarshney1986} as follows:
\begin{equation}
a_x=F_{k_{\rm f}}(\mathbf s_x)\triangleq\mathbf 1\big[\textstyle\sum_{j=1}^bs_{x,j}\ge k_{\rm f}\big]\label{eq:fusion}
\end{equation}
The late graph $G\triangleq([N],E,w)$ is simple and undirected, with weights $0<w_e\le\wmax$ and maximum degree $\Delta$. Any $\Delta$ is allowed; it bounds the class count, $C\le\Delta+1$ (Lemma~\ref{lem:colour}). The Laplacian is $\mathbf D-\mathbf W$ \cite{Shuman2013}. Here $\mathbf W$ is the weighted adjacency matrix, $\mathbf D\triangleq\mathrm{diag}(\mathbf W\mathbf 1)$ is the degree matrix and $\mathbf 1$ is the all-ones vector. Expanding the Laplacian gives:
\begin{equation}
\bL=\textstyle\sum_{\{u,v\}\in E}w_e(\ket u-\ket v)(\bra u-\bra v)\label{eq:laplacian}
\end{equation}
where $\ket x$, $x\in[N]$, is the computational basis state of the $d$-qubit address register (site $x$), and $e=\{u,v\}\in E$ has weight $w_e$. Edge colouring partitions $E$ into matchings $M_1,\dots,M_C$. Edge $e$ has colour $c_e$, with $c=1,\ldots,C$. The Laplacian splits as $\bL=\sum_{c=1}^C\bL_c$, with class Laplacians:
\begin{equation}
\bL_c\triangleq\textstyle\sum_{e\in M_c}w_e(\ket u-\ket v)(\bra u-\bra v)\label{eq:classlap}
\end{equation}
The planner that discloses $G$ supplies its colouring. Storing it as per-vertex $C$-bit colour masks would take $NC$ bits, linear in $N$. The quadratic form of \cite{Shuman2013} gives the normalized Dirichlet energy of the fused map as follows:
\begin{equation}
	\cE_G(\bz)=\bra{\psi_{\bz}}\bL\ket{\psi_{\bz}}=\bz^{\mathsf T}\bL\bz/N\label{eq:energydef}
\end{equation}
where $\ket{\psi_{\bz}}$ is the phase encoding \eqref{eq:psi} of Section~\ref{sec:loading}. The receiver performs one of the two tasks of Definition~\ref{def:tasks} on the energy \eqref{eq:energydef}: estimation, or a smoothness test between thresholds $\tau_0<\tau_1$. Its output, an estimate or a smooth/rough decision, is computed after epoch B from the register outcomes (Fig.~\ref{fig:overview}(b)).

\begin{definition}[Estimation and smoothness testing]\label{def:tasks}
An estimator outputs $\hat\cE$ with $|\hat\cE-\cE_G(\bz)|\le\varepsilon$ with probability at least $1-\delta$. A smoothness test decides between the hypotheses $\mathcal H_0$: $\cE_G(\bz)\le\tau_0$ (smooth) and $\mathcal H_1$: $\cE_G(\bz)\ge\tau_1$ (rough), with false-alarm and miss probabilities at most $\delta$.
The test is required only under this promise; energies strictly between the thresholds are excluded.
\end{definition}
A false alarm in Definition~\ref{def:tasks} decides rough for a smooth map, and a miss decides smooth for a rough map. The definition takes $\varepsilon>0$ and $0<\delta<1$ as given.

\begin{figure}[!t]
	\centering
	\includegraphics{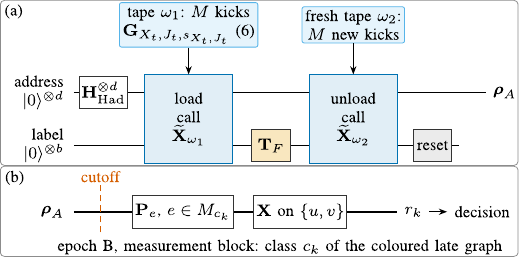}
	\caption{Circuit of one register of the quantum receiver. (a) Epoch A: the address register gets $\mathbf H_{\rm Had}^{\otimes d}$, and the load call applies signed kicks \eqref{eq:gate} of tape $\omega_1$ between label Hadamards and $\mathbf S_L^\dagger$ inside the call box. Fusion $\mathbf T_F$, an unload call with a fresh tape $\omega_2$, and label reset give the address state $\boldsymbol\rho_A$. The kicks are QOS phase updates \cite{ZhaoQOS}. The RR tapes are the two rounds ($M=Nb$, $\theta=\pi/2$). (b) Epoch B: register $k$ tests membership ($\mathbf P_e$) in an edge $e=\{u,v\}$ of its colour class $c_k$ after the cutoff. It measures the two-level subspace of $\ket u$ and $\ket v$ in the basis $(\ket u\pm\ket v)/\sqrt2$ ($\mathbf X$ on $\{u,v\}$). Its pair parity $r_k$ goes to the decision block (Section~\ref{sec:estimators}).}\label{fig:circuit}
\end{figure}

\subsection{Signed record kicks, calls, fusion, and reset}\label{sec:loading}
Each report carries one sub-bit. QOS writes a site's $b$ sub-bits into label qubits, where they are fused into a phase and erased. The fused sign remains as a phase of the address. Our table holds the $D=Nb$ raw sub-bits $s_{x,j}$, and a register holds a $d$-qubit address and a $b$-qubit label. A call is one block of $M$ reports applied to a register. Fig.~\ref{fig:circuit}(a) shows these steps for one register, described in turn below; $\mathbf H_L\triangleq\mathbf H_{\rm Had}^{\otimes b}$ applies the Hadamard gate $\mathbf H_{\rm Had} \triangleq\frac1{\sqrt2}\bigl[\begin{smallmatrix}1&1\\1&-1\end{smallmatrix}\bigr]$ and $\mathbf S_L\triangleq\mathbf S^{\otimes b}$ the phase gate $\mathbf S\triangleq\mathrm{diag}(1,i)$ to each label qubit.

\subsubsection{Load call}
Each register starts in $\ket{0^d,0^b}$ (Fig.~\ref{fig:circuit}). Address Hadamards $\mathbf H_{\rm Had}^{\otimes d}$ give $\ket{+}^{\otimes d}\ket{0^b}=N^{-1/2}\sum_x\ket{x,0^b}$, the start state of Section~\ref{sec:qos}. This uniform superposition of the basis states $\ket x$ of \eqref{eq:laplacian} is the load-call input. A report $(x,j,s)$ with $j\in\{1,\ldots,b\}$ and $s=s_{x,j}$ applies the signed kick at angle $\theta\triangleq\pi Nb/(2M)$ as follows:
\begin{equation}
\mathbf G_{x,j,s}\triangleq\exp\big(i\theta(-1)^s\ket x\!\bra x\otimes\ket1\!\bra1_j\big)\label{eq:gate}
\end{equation}
The label projector $\ket1\!\bra1_j$ acts on label qubit $j$, with identities on the others. Label basis vectors $\mathbf a,\mathbf c$ belong to $\{0,1\}^b$.
A block of $M$ reports applies $\mathbf U_\omega\triangleq\prod_t\mathbf G_{X_t,J_t,s_{X_t,J_t}}$. Its tape $\omega\triangleq((X_t,J_t))_{t=1}^M$ is the random sequence of (site, bit) indices of the $M$ reports arriving during one call, as in QOS (Section~\ref{sec:qos}). The tape is not a memory: the receiver applies each report as it arrives and discards it. The tapes $\omega_1$ and $\omega_2$ of the load and unload calls are the report sequences of two different blocks of arrivals. Substituting the mean counts $M/(Nb)$ defines the ideal target:
\begin{equation}
\mathbf U_{\rm kick}\ket{x,\mathbf a}\triangleq i^{|\mathbf a|}(-1)^{\mathbf a\cdot \mathbf s_x}\ket{x,\mathbf a},\label{eq:kick}
\end{equation}
with Hamming weight $|\mathbf a|\triangleq\sum_ja_j$, in general not $\E_\omega[\mathbf U_\omega]$. Under uniform pairwise-independent draws, signing makes each cell's phase-error variance independent of its bit.

Label Hadamards $\mathbf H_L$ and $\mathbf S_L^\dagger$ use the Bernstein--Vazirani phase-to-bit mechanism \cite{BernsteinVazirani1997} to convert the ideal $\mathbf U_{\rm kick}$ into the ideal bitwise exclusive OR (XOR) $\mathbf X_s\ket{x,\mathbf c}\triangleq\ket{x,\mathbf c\oplus \mathbf s_x}$. This is the common target of the load and unload calls. A call whose ideal action is $\mathbf X_s$ is an XOR call: it writes $\mathbf s_x$ into an empty label and removes it from a label that holds it. The realized call $\widetilde{\mathbf X}_\omega\triangleq\mathbf H_L\mathbf S_L^\dagger \mathbf U_\omega\mathbf H_L$ approximates it for tape $\omega$. The load call is $\widetilde{\mathbf X}_{\omega_1}$ with a first tape $\omega_1$. The unload call is $\widetilde{\mathbf X}_{\omega_2}$ with a fresh tape $\omega_2$. Ideally the load call writes the record into the label, $\ket{x,0^b}\mapsto\ket{x,\mathbf s_x}$.

QOS already includes indexed-bit subsampling \cite[Eqs.~(D91)--(D92)]{ZhaoQOS}; its complete-record updates are not directly available when each report carries one sub-bit (Appendix~\ref{app:prep}).

\subsubsection{Fusion}\label{sec:fusion}
The fusion rule acts through $\mathbf T_F\ket{\mathbf a}\triangleq(-1)^{F(\mathbf a)}\ket{\mathbf a}$; after the load call it ideally gives $z_x\ket{x,\mathbf s_x}$, the fused sign as a phase. Statuses are never reported.

\subsubsection{Unload call}
An ideal second XOR call uncomputes the record, leaving the fused sign as a phase:
\begin{equation}
\mathbf X_s\mathbf T_F\mathbf X_s\ket{x,0^b}=z_x\ket{x,0^b}\label{eq:phaseenc}
\end{equation}
so the ideal sequence applied to $N^{-1/2}\sum_x\ket{x,0^b}$ leaves the phase encoding of the fused map:
\begin{equation}
\ket{\psi_{\bz}}\triangleq N^{-1/2}\textstyle\sum_{x\in[N]}z_x\ket x\label{eq:psi}
\end{equation}
The approximate unload call $\widetilde{\mathbf X}_{\omega_2}$ uses a fresh, independent tape $\omega_2$. Replay would instead use retained reports, a data-dependent summary or compiled operation, or a replayable source, with the associated memory and dependence accounted for. The RR tapes $\omega_1$ and $\omega_2$ are the two rounds. The sources resend every bit in round~2, providing a replayable source with no receiver memory. CB uses a fresh tape $\omega_2$.

\subsubsection{Label reset}
The reset $\mathcal R(\boldsymbol\rho)\triangleq\operatorname{Tr}_L(\boldsymbol\rho)\otimes\ket{0^b}\!\bra{0^b}$ clears the label but preserves its address marginal. The resulting one-register address state is denoted $\boldsymbol\rho_A$; it crosses the cutoff, and epoch B measures it (Section~\ref{sec:readout}).

\subsubsection{Centred kicks and label check}\label{sec:check}
A CB variant replaces $\ket1\!\bra1_j$ in \eqref{eq:gate} by $\ket1\!\bra1_j-\frac12$, one extra phase gate on the address flag. It fuses with $\mathbf T_F'\ket{\mathbf a}\triangleq(-1)^{F(\mathbf a)+|\mathbf a|}\ket{\mathbf a}$ and measures the label instead of resetting it. The register is kept if the label reads $0^b$ and reprepared otherwise. Each call then errs, to first order, only by label bit flips, which the check detects; a kept register's error is second order (Theorem~\ref{thm:prep}(v)).

\subsection{Tape averaging and call error}\label{sec:tape}
CB gives each cell a random number of reports, so a call only approximates the ideal XOR $\mathbf X_s$. Every RR call is exact.
A call block averaged over its random, discarded tape performs the channel $\mathcal C_M$ of Appendix~\ref{app:prep}, as in \cite{ZhaoQOS}. The call error $e_b$ is its largest trace distance from the ideal XOR on this construction's inputs. It obeys:
\begin{equation}
	e_b\le\kappa_b v/2
\label{eq:singlecall}
\end{equation}
where $\kappa_b\triangleq(1+\sqrt{1+1/b})/2$ and $v\triangleq\pi^2Nb^2/(4M)$ is the phase-variance scale. Exact mean counts would give the ideal XOR. Random counts add phase errors with variance at most $v|\mathbf a|/b$. Appendix~\ref{app:prep} defines $\mathcal C_M$ and $e_b$ and proves the bound \eqref{eq:singlecall} in three steps.

\subsection{Source contracts, cutoff, and epoch B}\label{sec:contracts}
So that quantum and classical receivers see the same data, this subsection fixes the source contracts, the cutoff and the epoch-B data. The \emph{raw-report contract} $\mathsf C_{\rm R}$ covers RR and CB in Table~\ref{tab:scenarios}. Each of its reports $(x,j,s_{x,j})$ carries one sub-bit of a static table kept outside the receiver during epoch A. RR supplies every cell exactly once per call in two complete scans. CB supplies mutually independent call blocks of uniform, pairwise-independent cell draws (Theorem~\ref{thm:prep}). Exactly-once delivery is a property of the delivery service, not of acknowledgements alone; uniform draws idealize CB sources. The \emph{ordered-capture contract} $\mathsf C_{\rm O}$ (OC) supplies each whole record once, in a public order.

The \emph{cutoff} ends external old-value access, including old-data reports, archives, replay and remeasurement. Any old-data record, summary or program retained for epoch B counts as boundary memory. Before the cutoff, no usable position information reaches the receiver: the arrival order, timing and signal strength of reports do not depend on the late graph. Epoch B supplies each graph tuple $(u,v,w_e,c_e)$ once, in any order. The parameters $N$, $b$, $F$, $C$, $\wmax$, $\tau_0<\tau_1$, the class allocation and device addresses are public before epoch A. The planner holds $G$ and its colouring; this memory is counted for neither receiver, and the bounds of Section~\ref{sec:classical} grant the classical receiver the same epoch-B stream.

\subsection{Receiver memory and cost model}\label{sec:costmodel}
The \emph{boundary memory} is everything retained across the cutoff, including raw cells, labels, scratch, reports, seeds, flags, counters, control, and data-dependent programs. A quantum receiver's width is its peak number of logical qubits over both epochs; classical working storage has its own peak and boundary counts. A small quantum width alone does not bound controller, planner, or physical hardware resources. The Holevo bound \cite[Thm.~12.1]{NielsenChuang2010} limits what $Q$ boundary qubits add about the records, given the classical side information, to $Q$ bits under any later measurement. Throughout, logical qubits and their gates are ideal (error-corrected), and reports carry their bits without transmission errors under the contracts of Section~\ref{sec:contracts}.

We compare bits with logical qubits, counting classical control, as in \cite{LeGall2006} and \cite{Kallaugher2021}; QOS also distinguishes memory from processing time \cite{ZhaoQOS}. Comparable per-report operation rates are a modelling assumption. Gate work, source reports, and control are separate costs. Our online implementation applies reports as they arrive and discards them. Offline compilation instead introduces classical summaries and program storage \cite{Markidis2026}; these count while present, and any that survive the cutoff count as boundary memory.

\section{Graph-Spectral Layer}\label{sec:spectral}

This section shows which graph metrics the quantum receiver can obtain from its measurements, using graph signal processing only. After the cutoff a register never outputs $\bz$; it reveals only whether the signs at the two ends of one vertex pair of its colour class agree (Section~\ref{sec:readout}). A graph-spectral quantity is therefore available if it is a known constant plus a weighted sum of pair disagreements. Every quadratic spectral functional has this form (Section~\ref{sec:quadratic}). Its pairs split into colour classes of disjoint pairs, one class per register. The number of classes sets the register count (Section~\ref{sec:colour}).

\paragraph*{Graph frequencies and spectral summaries}
Laplacian eigenvalues $\lambda_k$ act as graph frequencies. The graph Fourier transform (GFT) splits $\bz$ into modes with normalized energies $p_k$, which form the map's graph periodogram. A spectral response $h(\lambda)$ weights these frequencies, and the functional $S_h=\sum_kh(\lambda_k)p_k$ of Section~\ref{sec:quadratic} covers the usual summaries of a graph signal. The response $h(\lambda)=\lambda$ gives the Dirichlet energy $\cE_G$, i.e.\ the mean frequency, and heat or diffusion weights $e^{-\tau\lambda}$, $\tau\ge0$, emphasize smooth content. Squared filter responses give a graph filter's output energy and a filter bank's band energies for estimating power spectral density (PSD) band by band. They also give spectral graph wavelet scale energies, showing how the map's energy spreads over scales. For a sensor map, these summaries describe how often neighbouring site decisions disagree and across which graph frequencies and scales.

The Laplacian eigenbasis gives $\bL=\mathbf U\boldsymbol\Lambda\mathbf U^{\mathsf T}$, with orthonormal eigenvector columns $\mathbf u_k$ and ordered eigenvalues $0=\lambda_1\le\cdots\le\lambda_N$. The GFT is $\hat\bz\triangleq\mathbf U^{\mathsf T}\bz$ \cite{Shuman2013}. Its normalized mode energies are $p_k\triangleq\hat z_k^2/N$, with $\sum_kp_k=1$. Let $W\triangleq\sum_ew_e$ and $D_e\triangleq\mathbf 1[a_u\ne a_v]$. The weighted degree is $\deg_w(x)\triangleq\sum_{e\ni x}w_e$. Lemma~\ref{lem:dirichlet} writes the energy \eqref{eq:energydef} in two further forms and bounds it. The spectral form $\sum_k\lambda_kp_k$ is the mean graph frequency: a smooth map puts its energy on low frequencies, and $\cE_G$ is the mean of the spectral distribution estimated in Section~\ref{sec:spectrum}. The edge form restates \eqref{eq:energyintro} with the disagreement indicators $D_e$; the pair-parity registers of Section~\ref{sec:quantum} measure it. The upper bound on $\lambda_N$ gives the spectral range used in Section~\ref{sec:spectrum}.
\begin{lemma}[Dirichlet energy of a fused sign map]\label{lem:dirichlet}
For every graph and fused sign map in the stated model, specializing the quadratic form in \cite{Shuman2013} gives the following:
\begin{equation}
\cE_G(\bz)=\textstyle\sum_{k=1}^N\lambda_kp_k=(4/N)\sum_{e\in E}w_eD_e\label{eq:dirichlet}
\end{equation}
and $0\le\cE_G\le\lambda_N\le2\max_x\deg_w(x)$.
\end{lemma}
A unit-weight perfect matching has $2$ as its only nonzero Laplacian eigenvalue. Its energy is $\cE_G=2q$, where $q$ is the matched disagreement fraction carried by the eigenvalue-$2$ band. A unit-weight $\Delta$-regular graph that factors into $\Delta$ perfect matchings has average class disagreement $\cE_G/(2\Delta)$.

\subsection{Quadratic spectral functionals}\label{sec:quadratic}
Using the spectral filter in \cite{Shuman2013}, define for a real response $h$:
\begin{equation}
S_h(\bz)\triangleq\textstyle\sum_{k=1}^Nh(\lambda_k)p_k=\bz^{\mathsf T}h(\bL)\bz/N\label{eq:Sh}
\end{equation}
where $h(\lambda)=\lambda$ gives $\cE_G$, $h(\lambda)=\lambda^j$, $j\ge0$, the spectral moments, and $e^{-\tau\lambda}$, with diffusion parameter $\tau\ge0$ in inverse graph-frequency units, a low-frequency weighting. Let $g_{{\rm bank},m}$ denote a filter-bank response indexed by $m$ and $g_{\rm wav}$ a wavelet response at scale $t_{\rm wav}>0$.
A polynomial of degree at most $K$ gives a $K$-hop finite impulse response (FIR) graph filter \cite{Shuman2013}, \cite{Sandryhaila2013}; a filter with response $h_{\rm f}$ has output energy $\|h_{\rm f}(\bL)\bz\|^2/N=S_{|h_{\rm f}|^2}$. Mode energies $p_k$ form a single-realization graph periodogram of $\bz$ \cite{Marques2017}. The energies $NS_{|h_{\rm f}|^2}$ with Chebyshev surrogates for nonpolynomial responses include two further summaries. The first are smooth filter-bank PSD estimates \cite{Marques2017}, the numerator of the single-realization Bartlett/Welch-type estimate, whose denominator $\|g_{{\rm bank},m}(\bL)\|_F^2$ does not depend on $\bz$ \cite{Perraudin2017}. The second are spectral graph wavelet scale energies $\|g_{\rm wav}(t_{\rm wav}\bL)\bz\|^2$ \cite{Hammond2011}. Theorem~\ref{thm:kernel} estimates them from pair parities.

Let $D_{uv}\triangleq\mathbf1[a_u\ne a_v]$ for any pair $u,v$ and $H_{uv}$ be entry $(u,v)$ of $\bH$. Lemma~\ref{lem:kernel} writes these functionals as sums over vertex pairs, which the registers of Section~\ref{sec:quantum} measure.
\begin{lemma}[Quadratic spectral functionals are pair functionals]\label{lem:kernel}
For real symmetric $\bH$ and $\bz\in\{\pm1\}^N$, $\bz^{\mathsf T}\bH\bz=\tr\bH+2\sum_{u<v}H_{uv}(1-2D_{uv})$; if $\bH=h(\bL)$ with polynomial degree $\deg h\le K$, then $H_{uv}=0$ beyond hop distance $K$.
\end{lemma}
\begin{proof}
Use $z_u^2=1$ and $z_uz_v=1-2D_{uv}$. Every nonzero product in $(\bL^j)_{uv}$ follows graph edges or diagonal stays, hence uses at most $j$ edges; $\bL^0=\mathbf I$.
\end{proof}

Polynomial responses are directly measurable: by Lemma~\ref{lem:kernel}, a response of degree at most $K$ involves only pairs within $K$ hops, which the colour classes of the $K$-hop graph cover (Section~\ref{sec:colour}). A smooth nonpolynomial response is replaced by its degree-$K$ truncation in the Chebyshev polynomials $T_k(y)\triangleq\cos(k\arccos y)$ on $[-1,1]$ \cite{Trefethen2013}; since $\sum_kp_k=1$, this changes $S_h$ by at most the largest truncation error at the eigenvalues. The same polynomials give the moments from which Section~\ref{sec:spectrum} reconstructs the spectral distribution of $\bz$.

\subsection{Colour classes}\label{sec:colour} A register measures pairs from one class of disjoint pairs, so the number of classes sets the register counts of Sections~\ref{sec:quantum} and \ref{sec:spectrum}. Tori are two-dimensional grids with wrap-around edges, used as the regular test graphs in Section~\ref{sec:numerics}. A rectangular torus has binary coordinate widths $d_1,d_2$. The notation $\mathbb Z_m$ denotes the integers modulo $m$.

\begin{lemma}[Colourings of kernel supports]\label{lem:colour}
Consider the simple off-diagonal support graph of a real symmetric $\bH$, with maximum degree $\Delta_{\bH}$. Vizing's theorem gives at most $\Delta_{\bH}+1$ edge-colour classes, and its proof finds such a colouring in polynomial time. A bipartite support admits $\Delta_{\bH}$ classes \cite[Thms.~17.2, 17.4]{BondyMurty2008}. Even simple cycles of length at least four admit $2$ classes; rectangular nearest-neighbour tori with both side lengths even and at least four admit $4$. For $K\in\mathbb Z_{\ge0}$, translation-invariant $K$-hop kernels admit $C_K\le2K$ on $\mathbb Z_{2^d}$ and $C_K\le2K(K+1)$ on $\mathbb Z_{2^{d_1}}\times\mathbb Z_{2^{d_2}}$, with $d,d_1,d_2\ge1$.
\end{lemma}

\section{Quantum Receiver and Memory Separation}\label{sec:quantum}
This section assembles the receiver from the loading of Section~\ref{sec:model} and the pair functionals of Section~\ref{sec:spectral}. The receiver holds $R$ address registers, ideally independent copies of \eqref{eq:psi}. They are loaded in epoch A and read by pair parities in epoch B (Figs.~\ref{fig:circuit}(a) and (b)). Their outcomes are averaged classically. These steps form the collection, measurement and decision blocks in Algorithm~\ref{alg:pipeline} and the dashed outline in Fig.~\ref{fig:overview}(c).

\begin{algorithm}[!t]
\caption{Quantum late-graph spectral estimation}\label{alg:pipeline}
\begin{algorithmic}[1]
\Require public $N=2^d$, $b$, $F$, $C$, $\wmax$, $\tau_0<\tau_1$, uniform classes $c_k$ (Section~\ref{sec:readout}); $R$ registers in $\ket{+}^{\otimes d}\ket{0^b}$ (Fig.~\ref{fig:circuit})
\algblockhead{Sensing}{Secs.~\ref{sec:fused}, \ref{sec:contracts}}
\State site $x\in[N]$ senses its record $\mathbf s_x\in\{0,1\}^b$
\State reports carry $(x,j,s_{x,j})$ (RR, CB) or $(x,\mathbf s_x)$ (OC)
\algblockhead{Epoch-A collection \textmd{(one branch)}}{Secs.~\ref{sec:loading}, \ref{sec:prep}}
\If{RR} \Comment{Thm.~\ref{thm:prep}(iii), shared}
\State $\mathbf H_L$; round~1 kicks \eqref{eq:gate}, $\theta=\pi/2$; $\mathbf S_L^\dagger$; $\mathbf H_L$
\State $\mathbf T_F$; $\mathbf H_L$; round~2 as round~1; $\mathbf S_L^\dagger$; $\mathbf H_L$; reset labels
\ElsIf{CB} \Comment{Thm.~\ref{thm:prep}(i) serial, (ii) shared}
\State load call $\widetilde{\mathbf X}_{\omega_1}$: $M$ random kicks, $\theta=\pi Nb/(2M)$
\State $\mathbf T_F$; unload call with a fresh block; reset labels
\State serial: per register, one label; shared: once, $R$ labels
\ElsIf{OC} \Comment{Rem.~\ref{rem:direct}}
\State per record: $(-1)^{F(\mathbf s_x)}$ on $\ket x$ in all registers; discard
\EndIf
\algblockhead{Cutoff: erasure}{Sec.~\ref{sec:contracts}}
\State sources erase; no old-value access; epoch B begins
\algblockhead{Localization}{Secs.~\ref{sec:fused}, \ref{sec:colour}}
\State bit-blind localizer: $G$ and classes $M_c$ \Comment{Lemma~\ref{lem:colour}}
\algblockhead{Epoch-B stream \textmd{(one route)}}{Secs.~\ref{sec:contracts}, \ref{sec:joint}}
\State energy: each coloured edge $(u,v,w_e,c_e)$ once, any order
\State spectrum: each coloured $K$-hop pair once, with $[T_j(\tilde\bL)]_{uv}$
\algblockhead{Measurement}{Sec.~\ref{sec:readout}}
\For{each arriving edge or pair $e$ of class $c_e$}
\For{each active register $k$ with $c_k=c_e$}
\State test $\mathbf P_e$; on a hit measure, release $(e,r_k)$  \hspace{-0.05in}  \Comment{Lem.~\ref{lem:outcome}}
\EndFor
\EndFor
\algblockhead{Decision}{Secs.~\ref{sec:estimators}, \ref{sec:joint}}
\State $\hat\cE=\frac1R\sum_k2Cw_{e_k}r_k$ ($0$ on a miss) \Comment{Thm.~\ref{thm:dirichlet}}
\State rough iff \rule{0pt}{2.6ex}$\hat\cE\ge(\tau_0+\tau_1)/2$
\State spectrum route: moments \eqref{eq:jointest}, distribution $\hat F_K$ \Comment{Thm.~\ref{thm:spectrum}}
\end{algorithmic}
\end{algorithm}

\subsection{Measurement: colour-class pair-parity readout}\label{sec:readout}
Each register reveals after the cutoff at most one edge of its colour class and whether its endpoint signs agree. The pair measurement adapts the quantum HM protocol \cite[Sec.~4]{BJK2004} from perfect matchings to supplied colour classes with a no-hit outcome. Register $k$ receives a class $c_k$ before epoch B, uniformly at random for Theorem~\ref{thm:dirichlet} or deterministically for the class-mean estimator of Section~\ref{sec:estimators}. Each tuple $(u,v,w_e,c_e)$ prompts every active register with $c_k=c_e$ to test coherently whether its address lies in $e$, using $\mathbf P_e\triangleq\ket u\!\bra u+\ket v\!\bra v$. Register $k$ hits edge $e_k\triangleq e$ if the test succeeds. It misses and reports nothing if no edge of its class contains its address. Only joint membership is measured, preserving endpoint coherence. On a hit, register $k$ measures the pair basis:
\begin{equation}
\ket{e,r}\triangleq(\ket u+(-1)^r\ket v)/\sqrt2,\qquad r\in\{0,1\},\label{eq:pairbasis}
\end{equation}
whose index $r$ is the parity, and is released with $(w_e,r_k)$. Since $\ket{e,1}\!\bra{e,1}=\frac12(\ket u-\ket v)(\bra u-\bra v)$, the following holds:
\begin{equation}
\textstyle\sum_{e\in M_c}w_e\ket{e,1}\!\bra{e,1}=\bL_c/2\label{eq:classeffect}
\end{equation}
Lemma~\ref{lem:outcome} gives the outcome law of one register of class $c$. Here $\Pr(e,r)$ is the probability of hitting edge $e\in M_c$ with pair parity $r$. The miss probability is $\Pr(\mathrm{miss})$.
\begin{lemma}[Outcome law]\label{lem:outcome}
For an ideal register of class $c$:
\begin{equation}
\Pr(e,r)=(z_u+(-1)^rz_v)^2/(2N),\qquad e\in M_c,\label{eq:outcome}
\end{equation}
and $\Pr(\mathrm{miss})=1-2|M_c|/N$; equivalently, a uniform latent site $X_{\rm site}$ covered by $M_c$ yields $(e,r)=(e(X_{\rm site}),D_{e(X_{\rm site})})$.
\end{lemma}
For any address state $\boldsymbol\rho$ and an independent uniform class, \eqref{eq:classlap} and \eqref{eq:classeffect} give the mean of the score $Y\triangleq2Cw_er$ of a register that hits edge $e$ with parity $r$ ($Y\triangleq0$ on a miss):
\begin{equation}
\E[Y\mid\boldsymbol\rho]=(1/C)\textstyle\sum_{c=1}^C2C\operatorname{Tr}(\boldsymbol\rho \, \bL_c/2)=\operatorname{Tr}(\boldsymbol\rho \, \bL)\label{eq:meanrho}
\end{equation}
It is $\cE_G(\bz)$ for an ideal register, and a prepared register has the bias $\operatorname{Tr}((\boldsymbol\rho-\ket{\psi_{\bz}}\!\bra{\psi_{\bz}})\bL)$, which the analytic curves of Section~\ref{sec:e2e} evaluate. Let $\delta_p>0$ be the allowed joint preparation distance and $\mathcal A$ any event of the complete readout. Equation~\eqref{eq:transfer} bounds each decision-probability change by $\delta_p$ when prepared copies replace ideal ones. Theorems~\ref{thm:dirichlet} and~\ref{thm:spectrum} and the certificate allowances of Appendix~\ref{app:cert} use this bound to transfer their ideal-register guarantees to the registers of Theorem~\ref{thm:prep}. The same bound applies to the event of Theorem~\ref{thm:kernel}(iii). Fix the graph, edge stream and decision rule, with class choices independent of the registers. The complete readout is then the same measurement for every joint register state. The measurement-event bound \cite[Thm.~9.1]{NielsenChuang2010} then gives the following for every joint state $\bar{\boldsymbol\rho}$ that satisfies $D_{\rm tr}(\bar{\boldsymbol\rho},\boldsymbol\psi_{\bz}^{\otimes R})\le\delta_p$, where $\boldsymbol\psi_{\bz}\triangleq\ket{\psi_{\bz}}\!\bra{\psi_{\bz}}$:
\begin{equation}
	\textstyle\big|\Pr_{\bar{\boldsymbol\rho}}(\mathcal A)-\Pr_{\boldsymbol\psi_{\bz}^{\otimes R}}(\mathcal A)\big|\le\delta_p\label{eq:transfer}
\end{equation}

\subsection{Decision: unbiased estimators}\label{sec:estimators}
The decision block of Algorithm~\ref{alg:pipeline} averages the register outcomes, each one random number whose mean is the target, into unbiased estimators. The following random-class bounds use independent ideal copies and independent uniform class choices. Concentration follows from Bernstein's inequality \cite[Cor.~2.11]{Boucheron2013}; deterministic allocation is treated separately.
Let $\delta_s\in(0,1)$ be the ideal sampling-failure allowance; Theorem~\ref{thm:dirichlet} averages the outcomes of Lemma~\ref{lem:outcome}.
\begin{theorem}[Dirichlet-energy estimator]\label{thm:dirichlet}
Let $Y_k\triangleq2Cw_{e_k}r_k$ on a hit, $Y_k\triangleq0$ on a miss, and $\hat\cE\triangleq R^{-1}\sum_{k=1}^RY_k$. Then (i) $\E[\hat\cE]=\cE_G(\bz)$; (ii) $0\le Y_k\le \, 2 \, C \, \wmax$ and $\E[Y_k^2]=(8 \,C/N)\sum_e w_e^2 D_e\le \, 2\,C \, \wmax\cE_G(\bz)$; (iii) with $\gamma\triangleq\tau_1-\tau_0$, alarming iff $\hat\cE\ge\tau_{\rm m}\triangleq(\tau_0+\tau_1)/2$ errs with probability at most $\delta_s$ under either hypothesis if the following holds:
\begin{equation}
R \ge 16 \, C\,\wmax(\tau_1+\gamma/6)\gamma^{-2}\ln(1/\delta_s);\label{eq:Rbern}
\end{equation}
(iv) $\Pr\{|\hat\cE-\cE_G|\le\varepsilon\}\ge1-\delta$ if $R\ge \,4 \, C \, \wmax(\cE_{\max}+\varepsilon/3)\varepsilon^{-2}\ln(2/\delta)$ for any $\cE_{\max}\ge\cE_G$. For the prepared state of Theorem~\ref{thm:prep}, each probability changes by at most $\delta_p$.
\end{theorem}

Let $H_e\triangleq H_{uv}$, $e=\{u,v\}$, and $\bH_{\rm off}\triangleq\bH-\mathrm{diag}((H_{xx})_{x=0}^{N-1})$. Theorem~\ref{thm:kernel} estimates the pair functionals.
\begin{theorem}[Kernel estimator]\label{thm:kernel}
Let $\bH$ be real symmetric with known trace and off-diagonal support coloured into $C_{\bH}\ge1$ classes, $Z_k\triangleq C_{\bH}H_{e_k}(-1)^{r_k}$ on a hit, $Z_k\triangleq0$ on a miss, $h_{\rm off}\triangleq\max_{u\ne v}|H_{uv}|$, and $\hat S\triangleq\tr(\bH)/N+R^{-1}\sum_kZ_k$. Then (i) $\E[\hat S]=\bz^{\mathsf T}\bH\bz/N$; (ii) $|Z_k|\le C_{\bH}h_{\rm off}$ and $\E[Z_k^2]=(C_{\bH}/N)\|\bH_{\rm off}\|_F^2$; (iii) for $\bH=h(\bL)$, with $\overline{h^2}\triangleq N^{-1}\sum_kh(\lambda_k)^2$, $\E[Z_k^2]\le C_{\bH}\overline{h^2}$, and $|\hat S-S_h|\le\varepsilon$ with probability at least $1-\delta$ if $R\ge(2C_{\bH}\overline{h^2}+\frac43C_{\bH}h_{\rm off}\varepsilon)\varepsilon^{-2}\ln(2/\delta)$.
\end{theorem}

The class-mean estimator is unbiased for fixed class counts $R_c$, and equal allocation $R_c=R/C$ has variance no larger than random classes (Appendix~\ref{app:est}). The certificate \eqref{eq:certificate} shows that $R=112$ registers so allocated suffice for false-alarm and miss probabilities below $0.027$ and $0.043$ on the hard family of Section~\ref{sec:classical} for $N\ge128$, for ideal copies only (Appendix~\ref{app:cert}).

A logical $\mathbf Z$ error of probability $p_{\rm Z}$ per address qubit leaves the hit law unchanged. Such errors, if independent across the address qubits, flip the parity of a hit on $e=\{u,v\}$ with probability $[1-(1-2p_{\rm Z})^{h_e}]/2$, where $h_e$ is the Hamming weight of $u\oplus v$. Gray addresses give $h_e=1$ on power-of-two cycles and tori, against $h_e\approx d/2$ for random ones.

\subsection{Epoch-A collection: register preparation}\label{sec:prep}\label{sec:resources}%
This subsection bounds the distance between the registers prepared under $\mathsf C_{\rm R}$ by the steps of Section~\ref{sec:loading} and ideal copies of $\ket{\psi_{\bz}}$. Let $L_N\triangleq(1+1/(2\sqrt N))^2$ and $\chi_{R,N}\triangleq(R+\sqrt{2RL_N+3R(R-1)})/4$; Theorem~\ref{thm:prep} gives sufficient report budgets for a target distance $\delta_p$.

\begin{theorem}[Preparation of $R$ copies]\label{thm:prep}
	Suppose the report blocks are mutually independent and random, with IID uniform or pairwise-independent uniform cell reports within each block. Then the tape-discarded averaged joint address state is within trace distance $\delta_p$ from $\ket{\psi_{\bz}}^{\otimes R}$ in either case: (i) $2R$ fresh blocks of $M_{\rm ser}\triangleq\lceil R\kappa_b\pi^2Nb^2/(4\delta_p)\rceil$ reports and one reused label (serial); or (ii) two blocks of $M_{\rm sh}\triangleq\lceil\chi_{R,N}\pi^2Nb^2/(2\delta_p)\rceil$ reports applied to all registers (shared). (iii) If instead every block is complete and scans every cell once, the state is the exact target ($2RNb$ or $2Nb$ reports). (iv) In each case the serial and shared widths are as follows:
	\begin{equation}
		Q_{\rm ser}=Rd+b+n_{\rm aux},\qquad Q_{\rm sh}=R(d+b)+n_{\rm aux}\label{eq:widths}
	\end{equation}
	with $n_{\rm aux}\triangleq\max\{d-1,b-2\}$. (v) For IID uniform reports and $v\le\frac15$, the centred kicks and label check of Section~\ref{sec:check} keep a register with probability at least $1-v$, and each kept register is within trace distance $v^2$ of $\ket{\psi_{\bz}}$.
\end{theorem}
The shared bound (ii) retains the cross-register dependence of common tapes. Part (v) gives, for $\delta_p\le R/25$, a sufficient serial budget of $\lceil\pi^2Nb^2\sqrt{R/\delta_p}/4\rceil$ reports per block. The $2R$ serial blocks of the kept registers total about $3.3\times10^5$ reports per raw bit at $R=112$, $b=4$ and $\delta_p=0.005$, compared with $5.2\times10^7$ from (i). RR full-step kicks load reversibly and exactly: each report with $s_{x,j}=1$ flips label bit $j$ of address $x$. RR therefore uses no QOS averaging, whereas CB's random repeated reports do. Table~\ref{tab:scenarios} identifies RR as case (iii), shared, with $Q_{\rm sh}=R(d+b)+d-1=113\log_2N+447$ at $R=112$, $b=4$ and $N\ge2^{b-1}$. For the same parameter setting, CB is case (i), serial, with $Q_{\rm ser}=113\log_2N+3$, or case (ii), shared, with $Q_{\rm sh}$. Its sufficient budgets at $\delta_p=0.005$ (the hard-family test) are about $5.2\times10^7$ (serial, $2RM_{\rm ser}/(Nb)$) and $6.0\times10^5$ (shared, $2M_{\rm sh}/(Nb)$) reports per raw bit, independent of $N$. OC uses the direct phase (Remark~\ref{rem:direct}). Table~\ref{tab:compare} lists both schedules. The factor $R$ in \eqref{eq:widths} counts the single-shot copies that the estimator averages; the accuracy target sets it, e.g., \eqref{eq:Rbern}, independently of $N$. All $R$ copies cross the cutoff, since the graph arrives after erasure; only the label and the ancillas are time-shared.

\begin{proposition}[Serial registers]\label{prop:serial}
	The serial schedule with $M$ reports per block gives independent registers, each within trace distance $D_1\le\kappa_bv$ of $\ket{\psi_{\bz}}$. The test of Theorem~\ref{thm:dirichlet}(iii) errs with probability at most $\delta_s$ under either hypothesis if $\beta\triangleq\lambda_ND_1<\gamma/2$ and the following holds:
	\begin{equation}
		R\ge16\,C\,\wmax(\tau_1+\gamma/6)(\gamma-2\beta)^{-2}\ln(1/\delta_s)\label{eq:Rserial}
	\end{equation}
\end{proposition}
Proposition~\ref{prop:serial} bounds the bias of each register instead of the joint distance $\delta_p$: $\lceil\kappa_b\pi^2Nb^2\lambda_N/(4\beta)\rceil$ reports per block suffice, i.e., $R\kappa_b\pi^2b\lambda_N/(2\beta)$ per raw bit. Lemma~\ref{lem:dirichlet} and $\deg_w(x)\le C\wmax$ give $\lambda_N\le2C\wmax$ before the graph is known.

\begin{remark}[Site-contiguous direct phase]\label{rem:direct}
	Under $\mathsf C_{\rm O}$, buffering one record classically and applying $(-1)^{F(\mathbf s_x)}$ with an address flag prepares the target exactly:
	\begin{equation}
		\textstyle\prod_{x=0}^{N-1}\big(\mathbf I-2F(\mathbf s_x)\ket x\!\bra x\big)N^{-1/2}\sum_x\ket x=\ket{\psi_{\bz}}\label{eq:directphase}
	\end{equation}
	This complete scan is the deterministic case of \cite[Eqs.~(2)--(4)]{ZhaoQOS}, with each site visited once at angle $\pi$, as in the streaming protocol of \cite[Sec.~1.4]{GKKRW}. It uses no label or unload call. All registers are updated before the record is discarded.
\end{remark}
\begin{remark}[Departures from the contracts]\label{rem:robust}
	If the two RR rounds are inconsistent at $m$ sites (lost, duplicated, flipped or changed sub-bits), the mean score \eqref{eq:meanrho} shifts by at most $4m\max_x\deg_w(x)/N$, and by at most half of that when each such site receives two complete but different records. Suppose CB has unequal cell probabilities $p_c\ge p_{\min}$ that the controller computes from the cell index. Kicks at angle $\pi/(2Mp_c)$ preserve Theorem~\ref{thm:prep}(i)--(ii), with $Nb^2$ replaced by $b/p_{\min}$ (Appendix~\ref{app:prep}). QOS also extends to unknown, non-uniform and correlated sampling \cite[Thm.~D.16, Lemma~D.18]{ZhaoQOS}.
\end{remark}
The error allowances of the certified test and its widths up to $N=2^{40}$ are in Appendix~\ref{app:cert}.

\subsection{Classical lower bound and memory separation}\label{sec:classical}
 The \emph{hard family} embeds $\alpha$PM \cite{GKKRW} into a connected late graph on $N=4n$ sites with $C=4$ unit-weight classes, maximum degree $4$ and $\cE_G=3+\beta-1/n$, so the energy reveals the hidden answer $\beta$. Appendix~\ref{app:classical} constructs this family and turns any classical receiver into a one-way protocol whose message is its boundary memory. It obeys the randomized one-way lower bound of \cite[Thm.~2]{GKKRW} at $\alpha=1/4$. This uses the Section~\ref{sec:contracts} condition that no usable position information reaches the receiver before the cutoff.
\begin{theorem}[Classical boundary memory]\label{thm:classical}
Fix any nonconstant fusion rule. Every classical receiver deciding $\cE_G\le\tau_0$ versus $\cE_G\ge\tau_1$ with error at most $\frac13$ on all admitted instances needs $\Omega(\sqrt N)$ boundary bits in each family: (a) unit-weight perfect matchings with fixed $0\le\tau_0<\tau_1\le1$; (b) the hard family of Section~\ref{sec:classical} with $\tau_0=3-1/n$ and $\tau_1=4-1/n$ ($C=4$, $\gamma=1$).
\end{theorem}
A \emph{site cache} keeps the records of $s_{\rm c}$ sites sampled by a hash before the cutoff. It then rescales disagreements on edges with both ends kept. The estimator is unbiased. A cache of $s_{\rm c}=O(\sqrt N)$ sites, or $O(\sqrt N\log N)$ bits, suffices at fixed $b$, $C$, $\wmax$, $\gamma$ and $\delta_s$ (Appendix~\ref{app:classical}), nearly matching this bound.
\begin{corollary}[Memory separation]\label{cor:separation}
Assume the strict cutoff, nonconstant fusion, fixed $C$, $\wmax$ and $\gamma$, and a total error target $0<\delta\le1/3$ that contains fixed positive sampling and preparation allowances and every other applicable failure allowance. Then (a) the smoothness test of Theorem~\ref{thm:dirichlet} uses $R=O(1)$ registers. Its logical width is $O(\log N)$ for fixed $b$ or $b=O(\log N)$. Its classical control uses $O(\log N)$ bits if its class/address fields, fusion program, counters, precision and queues each fit in $O(\log N)$ bits. (b) Every classical receiver needs $\Omega(\sqrt N)$ boundary bits on the hard family at the thresholds of Theorem~\ref{thm:classical}(b).
\end{corollary}
\begin{proof}
Combine Theorems~\ref{thm:dirichlet}, \ref{thm:prep} and~\ref{thm:classical}.
\end{proof}
The separation rests on late, unpredictable graphs. Disclosing $G$ before epoch A lets the receiver sample $O(1)$ edges and keep their endpoint records. This estimates the energy at fixed accuracy with $O(\log N)$ record bits. Sampling $O(1)$ patches of a promised nearest-neighbour torus does the same (Appendix~\ref{app:classical}). Reloading one register after the graph is known turns any instance into this early-graph case. A known backbone can keep the bound if its cross edges arrive late. A \emph{fixed ring} has banks $\mathsf L,\mathsf R$ of $N/2$ sites forming public cycles of weight-$1/4$ edges. Edge parity colours them into two classes $M_0,M_1$. The late cross edges are weight-one rungs $\{(\mathsf L,i),(\mathsf R,\varpi(i))\}$, as on a ladder. They pair the banks by a permutation $\varpi$ that arrives in epoch B. Equation~\eqref{eq:dirichlet} gives this graph $G_\varpi$ energy $\cE_\varpi\triangleq\cE_{G_\varpi}=D_{\rm cyc}/N+2q_\varpi$, with $D_{\rm cyc}$ disagreeing cycle edges and a fraction $q_\varpi$ of disagreeing rungs. Theorem~\ref{thm:ring} extends Theorem~\ref{thm:classical} to this family; its admitted record domain is the set of records that the source may deliver.
\begin{theorem}[Classical memory with a fixed ring]\label{thm:ring}
	Fix nonconstant fusion on the admitted record domain and the strict old-value cutoff, under either source contract. For $N=16n$ with $n\ge4$ a power of two, every randomized classical receiver deciding $\cE_\varpi\le\frac12$ versus $\cE_\varpi\ge1$ with error at most $\frac13$ on all admitted instances retains $\Omega(\sqrt N)$ boundary bits in the worst case, although only the rungs are late.
\end{theorem}
On this family the quantum receiver uses $C=3$ classes, and \eqref{eq:Rbern} gives $R\ge208\ln(1/\delta_s)$ ideal registers.

\subsection{Computational complexity}\label{sec:complexity}
Table~\ref{tab:gates} of Appendix~\ref{app:runtime} lists the gate and classical operation counts for fixed $b$. Each kick \eqref{eq:gate} uses $d-1$ logical ANDs and $O(\log N)$ single- and two-qubit Clifford gates within the widths \eqref{eq:widths}. Each membership test uses $2(d-1)$ ANDs. The classical controller uses $O(1)$ word operations per kick, edge and test. RR applies $O(RNb\log N)$ gates in epoch A because each of its $2Nb$ reports kicks all $R$ registers. CB applies $O(R^2Nb^2\log N/\delta_p)$ gates in either schedule of Theorem~\ref{thm:prep}, compared with $O(RN)$ for the OC direct phase in compatible address order (Remark~\ref{rem:direct}). Per raw bit, RR applies $2R$ kicks and CB applies about $5.2\times10^7$ (serial) or $6.8\times10^7$ (shared) at $R=112$, $b=4$ and $\delta_p=0.005$. Each register tests at most the $N/2$ edges of its class in epoch B. The total is $O(RN\log N)$ gates. The separation concerns memory, not time. One controlled-controlled-$\mathbf Z$ factory per register keeps up with the average OC work for streams S1--S3 of Section~\ref{sec:massive} if its output rate is $25$ to $2.5\times10^5$ times faster than an illustrative surface-code benchmark. RR multiplies these factors by $2b(d-1)$.

\section{Late-Graph Spectral Estimation}\label{sec:spectrum}
Dirichlet energy is the mean of the fused map's energy distribution over graph frequencies; this section estimates the distribution from pair-parity Chebyshev moments. The weights $p_k=|\mathbf u_k^{\mathsf T}\bz|^2/N$ of Section~\ref{sec:spectral} define the signal-weighted spectral measure $\rho_{\bz}$. This parallels the local density of states \cite[Eq.~(112)]{Weisse2006}, with the site state replaced by $\ket{\psi_{\bz}}$ and unit total mass. The measure and its distribution function $F_{\bz}$ are:
\begin{equation}
\rho_{\bz}(\lambda)\triangleq\textstyle\sum_{k=1}^Np_k\,\delta_{\rm D}(\lambda-\lambda_k);\quad F_{\bz}(\lambda)\triangleq\sum_{k:\,\lambda_k\le\lambda}p_k,\label{eq:spectrum}
\end{equation}
where $\delta_{\rm D}$ is the Dirac delta function. Repeated eigenvalues aggregate their eigenspace weights, independently of the eigenbasis chosen. The measure spreads the map's squared magnitude over graph frequencies. It is a distribution for one fused map, without a stationarity assumption; disclosure follows the raw-data cutoff, whereas the process-PSD estimators of \cite{Marques2017} and \cite{Perraudin2017} use the graph in their transforms or filters.

The projective measurement rule \cite{NielsenChuang2010} identifies $p_k=|\langle\mathbf u_k|\psi_{\bz}\rangle|^2$. The receiver instead estimates polynomial expectation values from pair parities. Specializing the matrix-element moments of the kernel polynomial method (KPM) \cite{Weisse2006} to this state yields the moments used below for finite-resolution reconstruction, without eigenbasis readout.

\subsection{Chebyshev moments and the kernel polynomial method}\label{sec:kpm}
This subsection defines the Chebyshev moments of the fused map, the averages of the Chebyshev polynomials $T_j$ over its rescaled spectrum, and the KPM approximation of $F_{\bz}$ from them. Since $T_j$ lives on $[-1,1]$, the spectrum is first rescaled \cite{Weisse2006}. A Chebyshev series of $\rho_{\bz}$ truncated after $T_K$ oscillates (Gibbs); KPM multiplies the $j$th moment by a Jackson coefficient $g_j\in[0,1]$ \cite[Eq.~(71)]{Weisse2006}, which suppresses the oscillations and keeps the damped density nonnegative.
\begin{definition}[Chebyshev moments and KPM distribution]\label{def:kpm}
Let $\bar\lambda\ge\lambda_N$, $\bar\lambda>0$ (e.g., $2\max_x\deg_w(x)$) and $0<\epsilon_{\rm c}<2$. Define $a_\lambda\triangleq\bar\lambda/(2-\epsilon_{\rm c})$, $\tilde\bL\triangleq(\bL-\tfrac{\bar\lambda}2\mathbf I)/a_\lambda$ and $y(\lambda)\triangleq(\lambda-\bar\lambda/2)/a_\lambda$. This map sends the eigenvalues $\lambda_k$ of $\bL$ to $y_k\triangleq y(\lambda_k)\in(-1,1)$. Let $K\ge1$ be an integer. The Chebyshev moments of $\rho_{\bz}$ are as follows:
\begin{equation}
	\mu_j\triangleq\textstyle\sum_kp_kT_j(y_k)=\bz^{\mathsf T}T_j(\tilde\bL)\bz/N=\bra{\psi_{\bz}}T_j(\tilde\bL)\ket{\psi_{\bz}},\label{eq:chebmom}
\end{equation}
$j=0,\dots,K$, and the Jackson-damped KPM distribution, the integral of the damped density \cite{Weisse2006} from $-1$ to $y=\cos\varphi$, is:
\begin{equation}
	\bar F_K(y)\triangleq g_0(1-\varphi/\pi)-(2/\pi)\textstyle\sum_{j=1}^Kg_j\mu_j(\sin j\varphi)/j
	\label{eq:kpmcdf}
\end{equation}
for $-1\le y\le1$ and $\varphi\in[0,\pi]$, zero below $-1$ and one above $1$, read in graph frequency at $y=y(\lambda)$.
\end{definition}
Here $\mu_0=1$ and $|\mu_j|\le1$; the moments are quadratic forms of $j$-hop FIR graph filters (Lemma~\ref{lem:kernel}). The distribution $\bar F_K$ is a smoothed approximation of $F_{\bz}$ rebuilt from $\mu_0,\dots,\mu_K$. Its support is $[\bar\lambda/2-a_\lambda,\bar\lambda/2+a_\lambda]$ in graph-frequency units. Appendix~\ref{app:spectrum} gives $g_j$, the integration and the line widths, which describe how far damping spreads each eigenvalue. The accuracy of $\bar F_K$ is measured by the Wasserstein distance, the area between two distribution functions, i.e., how far spectral mass must move along the frequency axis to turn one into the other. For the eigenvalue density of a graph, a Jackson-damped reconstruction with closely related coefficients has Wasserstein error bounds that shrink as $K$ grows \cite[Eq.~(12), Lemma~A.3]{BKM2022}; Theorem~\ref{thm:spectrum} proves such bounds for $\rho_{\bz}$ when the moments come from pair parities.

\subsection{Joint moment estimation from pair parities}\label{sec:joint}
Each Chebyshev moment is a quadratic functional of Section~\ref{sec:quadratic}, so the same registers estimate all $K$ moments. Spectral estimation uses a different supplied stream from that of Section~\ref{sec:contracts}: each $K$-hop pair carries its class and polynomial coefficients. Colour the $K$-hop graph into $C_K\ge1$ matchings (Lemma~\ref{lem:colour}), allowing empty classes. Use independent ideal registers and independent uniform class choices. On a hit of $\{u_k,v_k\}$ with outcome $r_k$, register $k$ contributes $Y_{k,j}\triangleq C_K[T_j(\tilde\bL)]_{u_kv_k}(-1)^{r_k}$ to every moment $j$ at once:
\begin{equation}
\hat\mu_j\triangleq\operatorname{tr}T_j(\tilde\bL)/N+R^{-1}\textstyle\sum_{k=1}^RY_{k,j},\label{eq:jointest}
\end{equation}
for $1\le j\le K$, with $Y_{k,j}\triangleq0$ on a miss and $\hat\mu_0\triangleq\mu_0=1$. Let $F_K^{\rm raw}$ be \eqref{eq:kpmcdf} with these estimates. Clip it to $[0,1]$ and increasingly rearrange it with respect to uniform measure on $[-1,1]$. Take the right-continuous version, extended by zero below $-1$ and one at and above $1$. This defines the estimated distribution function $\hat F_K$, read in graph frequency at $y=y(\lambda)$, as is $\bar F_K$. Sampled moments alone need not define a positive measure. Distances use the uniform norm and, for two distribution functions $\mathcal F_1,\mathcal F_2$, the Wasserstein distance:
\begin{equation}
W_1(\mathcal F_1,\mathcal F_2)\triangleq\textstyle\int_{\mathbb R}|\mathcal F_1(\lambda)-\mathcal F_2(\lambda)|\,d\lambda\label{eq:wasserstein}
\end{equation}
in graph-frequency units. Theorem~\ref{thm:spectrum} bounds the moment and distribution errors of this estimator.

\begin{theorem}[Late-graph spectral estimation]\label{thm:spectrum}
Use the preceding independent ideal experiment, integer $K\ge1$, $\varepsilon>0$ and $0<\delta<1$. Let the planner stream each $K$-hop pair once in epoch B, with its class and coefficients $[T_j(\tilde\bL)]_{uv}$, $j\le K$. Let it also disclose the positive bound $\bar\lambda\ge\lambda_N$ and exact normalized traces. Let an upper bound on $C_K$ be public before epoch A, and let the register count $R$ satisfy:
\begin{equation}
R\ge(2C_K+4C_K\varepsilon/3)\varepsilon^{-2}\ln(2K/\delta).
\label{eq:Rspec}
\end{equation}
Then, with probability at least $1-\delta$: (i) $|\hat\mu_j-\mu_j|\le\varepsilon$ for $j=1,\dots,K$; (ii) $\sup_\lambda|\hat F_K-\bar F_K|\le\frac2\pi H_K\varepsilon$ with $H_K\triangleq\sum_{j\le K}\frac1j\le1+\ln K$; and (iii) $W_1(\hat F_K,F_{\bz})\le a_\lambda\big(2\sin\frac{\pi}{2K+4}+\frac4\pi H_K\varepsilon\big)$. Moreover, (iv) the receiver has width $Q_{\rm ser}$ (serial) or $Q_{\rm sh}$ (shared) of \eqref{eq:widths}, $O(\log N)$ for $b=O(\log N)$ and fixed $K$, $\varepsilon$, $\delta$, and $C_K$, and (v) prepared registers change these probabilities by at most $\delta_p$.
\end{theorem}
The count \eqref{eq:Rspec} follows from Bernstein\textquotesingle s inequality \cite{Boucheron2013} and a union bound. The public bound serves as $C_K$ in the class choices, the scores $Y_{k,j}$ and \eqref{eq:Rspec}; unused classes stay empty. QRAM, Store all and this work in Table~\ref{tab:compare} also give the spectral distribution. HM in \cite{GKKRW,KPV2024,KPV2025} lacks this estimator. The estimator's readout applies to any loader preparing the same phase states. The classical planner supplies graph-dependent coefficients. Its work is counted outside this receiver-stage theorem. Coefficient and trace errors shift each moment by at most their size, with coefficient errors multiplied by $C_K$. Original edges alone do not supply this interface.

The estimate describes a finite-resolution distribution, not the exact list of eigenvalues and coefficients. The resolution term $2a_\lambda\sin\frac{\pi}{2K+4}\approx\pi\bar\lambda/(2K)$, for large $K$ and small $\epsilon_{\rm c}$, decreases with $K$. The class count, and hence the register budget, can increase. Lemma~\ref{lem:colour} makes lattice budgets polynomial in $K$. Budgets can be much larger on rapidly growing graphs (Appendix~\ref{app:spectrum}). Once the $K$-hop neighbourhoods fill the graph, $C_K$ reaches $N-1$ and \eqref{eq:Rspec} grows with $N$, outside the fixed-$C_K$ regime of (iv). Corollary~\ref{cor:specclassical} extends the classical lower bound of Theorem~\ref{thm:classical} to the spectral distribution through its mean; hardness beyond the mean is open.

\begin{corollary}[Classical cost of spectral estimation]\label{cor:specclassical}
Under the assumptions of Theorem~\ref{thm:classical}(b), every classical receiver that outputs, with probability at least $\frac23$ on every admitted instance, a distribution function within a Wasserstein distance $\varepsilon_{\rm W}<\frac12$ of $F_{\bz}$ needs $\Omega(\sqrt N)$ boundary bits in the worst case.
\end{corollary}

By Theorem~\ref{thm:spectrum}(ii), $1-\hat F_K(\lambda_{\rm c})$ estimates the Jackson-damped spectral mass $1-\bar F_K(\lambda_{\rm c})$ above a cutoff frequency $\lambda_{\rm c}$ within $\frac2\pi H_K\varepsilon$. This estimate uses only the $C_K$ classes, whereas an ideal high-pass projector can have dense support.

\section{Simulations}\label{sec:numerics}
Table~\ref{tab:simsum} gives the parameters, the simulation type and the qubits of one simulated register for each simulated figure. Ideal-register experiments (exact in RR, Theorem~\ref{thm:prep}(iii)) sample the outcome law of Lemma~\ref{lem:outcome}, with the stated kernel modifications. The end-to-end runs of Sections~\ref{sec:e2e} and~\ref{sec:sensor} evolve each register's address-and-label state vector of $N2^b$ complex amplitudes under sampled report tapes. They draw readouts from exact Born probabilities. The gate-level runs of Section~\ref{sec:gatelevel} use only gates and measurements. Sensor maps use $b=4$ noisy one-bit looks (dimensionless Gaussian noise standard deviation $\sigma=0.6$) at a standardized heat-diffused field, independent per-site fault probability $0.02$, fair replacement bits and three-of-four fusion. The hard family uses the records of Appendix~\ref{app:classical}. The tori and the sensor graph validate preparation and estimation of graph functionals, not fault classification; the worst-case memory separation is proved on the hard family.

\begin{table}[!t]
\caption{Simulations: parameters, type and qubits}\label{tab:simsum}
\centering\scriptsize\setlength{\tabcolsep}{1pt}\renewcommand{\arraystretch}{1.0}\setlength{\extrarowheight}{2pt}
\begin{tabularx}{\columnwidth}{|c|>{\raggedright\arraybackslash}X|>{\raggedright\arraybackslash}p{0.22\columnwidth}|>{\raggedright\arraybackslash}p{0.25\columnwidth}|}\hline
Fig. & Parameters & Simulation & Qubits\\\hline
\ref{fig:memory}(a) & Hard family, $N=2^{10}$--$2^{18}$, $\delta_s=0.05$ & Monte Carlo cache, exact counts & --\\\hline
\ref{fig:memory}(b) & Streams S1--S4, OC, $b=4$, $R=112$ & Extrapolation & --\\\hline
\ref{fig:e2e} & $8\times8$ torus, $R=256$, $M=10^3$--$10^6$ & State vector & $d+b=10$\\\hline
\ref{fig:scaling} & Tori, $N=2^{10}$--$2^{20}$, $R=256$, $\kappa_bv=0.05,0.01$ & State vector & \mbox{$d+b=14$--$24$}\\\hline
\ref{fig:sensor}(a) & Sensor graph, $N=1024$, $C=17$, $R=256$, $M=5\times10^4$--$5\times10^6$ & State vector & $d+b=14$\\\hline
\ref{fig:sensor}(b) & Sensor graph, $K=4$--$64$, $R=10^5,10^6$ & Outcome law & --\\\hline
\ref{fig:gatelevel} & Tori, $C=4$, $N=2^6$--$2^{19}$, $R=16$--$2048$, $\kappa_bv=0.05,0.01$ & Gate level (Qiskit Aer, dedicated) & \mbox{$d+b+1=11$--$24$}\\\hline
\end{tabularx}
\end{table}

\subsection{Memory at fixed error and massive streams}\label{sec:massive}
\begin{figure}[!t]
\centering\includegraphics{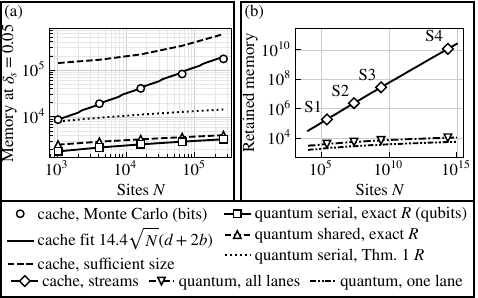}
\caption{Classical record bits and quantum-receiver logical qubits versus $N$ (logarithmic axes). (a) Connected hard family at a $5\%$ per-hypothesis sampling-error target. Cache sizes are interpolated from Monte Carlo trials: three instances per hypothesis, $300$ cache draws each. The plot shows their median fit $14.4\sqrt N(d+2b)$ and the sufficient size of Appendix~\ref{app:classical}. Ideal quantum widths use finite-search binomial budgets (serial, shared) and the Bernstein bound. (b) Extrapolated cache bits and logical widths for streams S1--S4 of Section~\ref{sec:massive} under OC. The quantum receiver uses the direct-phase branch (Remark~\ref{rem:direct}), with all registers active or one preparation lane, at the $R=112$ certificate.}\label{fig:memory}
\end{figure}

On the hard family for $N=2^{10}$--$2^{18}$ (Fig.~\ref{fig:memory}(a)), every random-class register count from $R=182$--$184$ on suffices for $5\%$ error. The Bernstein count \eqref{eq:Rbern} is $4.3$--$4.4$ times as large. The site cache reaches this error at an interpolated $s_{\rm c}\approx13.0$--$15.2\sqrt N$ sites, or $8.7\times10^3$ to $1.7\times10^5$ bits. The comparison at $N=2^{14}$ is $2.6\times10^3$ qubits against $4.1\times10^4$ bits. Storing all $Nb$ bits takes less than the cache up to $N=2^{12}$. The fitted cache at $N=2^{40}$ holds $7.3\times10^8$ bits ($91$~MB), against $4{,}523$ logical qubits at $R=112$ (Appendix~\ref{app:cert}). The interpolated sizes and their fit carry unquantified uncertainty.

\begin{figure}[!t]
\centering\includegraphics{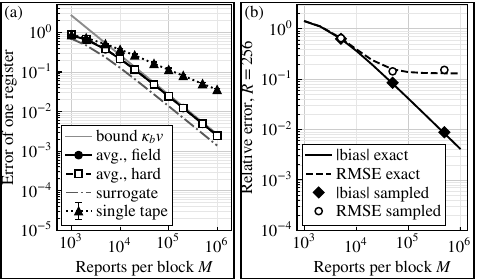}
\caption{End-to-end CB simulations with the Boolean-sketch code of \cite{ZhaoQOS} on an $8\times8$ unit-weight torus (heat parameter $\tau=2$), versus reports per block $M$ (logarithmic axes). (a) One-register error: analytic tape-averaged trace distances for this map (field) and an $N=64$ hard-family input (hard), the bound $\kappa_bv$, and the surrogate. The surrogate is an unnormalized mean-cell vector diagnostic in Euclidean distance. Fixed-realization errors for this map (single tape) show mean $\pm$ standard deviation over $40$ load/unload tape pairs, each of two $M$-report blocks. (b) Absolute relative bias and relative root-mean-square error (RMSE) of $\hat\cE$ for $R=256$: analytic averages (``exact'', lines) and sampled pipelines (markers, $40$ estimates per $M$).}\label{fig:e2e}
\end{figure}

Fig.~\ref{fig:memory}(b) extrapolates cache bits and logical qubits to four OC streams of $b=4$-bit records, one site per record. S1--S3 use one, $10^2$ and $10^4$ sensors at $1$\,Mb/s over one second; S4 uses $10^4$ sensors over one day. The reduction grows as $\sqrt N/R$. This extrapolation uses no new trials, inherits the fit uncertainty, and is not a physical-memory benchmark.

\subsection{End-to-end CB pipeline}\label{sec:e2e}
\begin{figure}[!t]
\centering\includegraphics{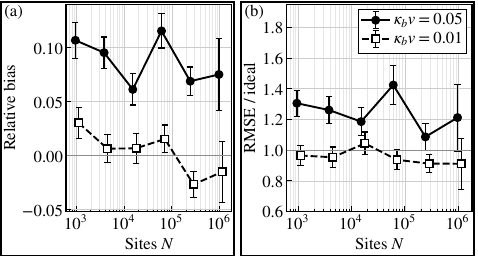}
\caption{End-to-end scaling on unit-weight tori versus the number of sites $N=2^{10}$--$2^{20}$ (logarithmic axis; $\tau=6$, $R=256$). Each size and preparation target uses $100$ estimates at $N\le2^{18}$ and $20$ at $N=2^{20}$, with one sensor map per size. The targets $\kappa_bv=0.05$ (filled) and $0.01$ (open) are offset in $N$ for legibility. (a) Relative bias of $\hat\cE$ and (b) RMSE divided by the analytic ideal-register RMSE, with one estimated standard error.}\label{fig:scaling}
\end{figure}

 Figs.~\ref{fig:e2e} and~\ref{fig:scaling} simulate the end-to-end CB pipeline with sampled tapes: Algorithm~\ref{alg:pipeline} with the calls of Fig.~\ref{fig:circuit}. The sampler and Boolean phase-oracle sketch of the QOS reference code \cite{ZhaoQOS} implement its tapes and kicks \eqref{eq:gate} in Fig.~\ref{fig:e2e}. The $8\times8$ torus has $\cE_G=1.5$. Fig.~\ref{fig:e2e}(a) shows tape-averaged preparation error decaying as $1/M$, at $0.80$--$0.92$ of $\kappa_bv$ for $M\ge10^4$. The mean fixed-realization error falls roughly as $M^{-1/2}$. Fig.~\ref{fig:e2e}(b) shows the exact-average relative bias of $\hat\cE$, the bias of \eqref{eq:meanrho} relative to $\cE_G$. It falls from $0.63$ to $0.008$ as $M$ grows from $5\times10^3$ to $5\times10^5$, consistent with sampled estimates. Register counts of $562$ to $199$ suffice on the hard family for $5\%$ error at $M=10^4$ to $10^5$, against $185$ for ideal registers. The sampled pipeline at $N=1024$ ($\cE_G=1.594$, $\kappa_bv=0.05$) returns $1.71\pm0.06$ (mean of $12$ estimates) with $R=256$. It uses $4.4\times10^8$ reports per estimate, against $2.1\times10^6$ for a serial scan ($2RNb$ at $R=256$; the shared RR count $2Nb$ is $8{,}192$). The memory advantage costs reports and gate work. The $R=112$ test at $\delta_p=0.005$ has sufficient budgets from Theorem~\ref{thm:prep} of $2.1\times10^{11}$ serial or $2.5\times10^9$ shared reports at this $N$. These budgets are linear in $N$ and not proved necessary.

Fixing $b$, $R$ and $\kappa_bv$ makes the preparation bound of Theorem~\ref{thm:prep} independent of $N$. Tori up to $N=2^{20}$ sites show no resolved bias growth with $N$ (Fig.~\ref{fig:scaling}(a)). The bias is $+6\%$ to $+12\%$ at $\kappa_bv=0.05$ and $-3\%$ to $+3\%$ at $0.01$. The RMSE stays within $0.91$--$1.42$ times the ideal-register RMSE (Fig.~\ref{fig:scaling}(b)), with standard errors up to $\figSixRmseSE$. The serial width grows only from $2{,}573$ to $5{,}143$ logical qubits.

\subsection{A sensor-network graph}\label{sec:sensor}
\begin{figure}[!t]
\centering\includegraphics{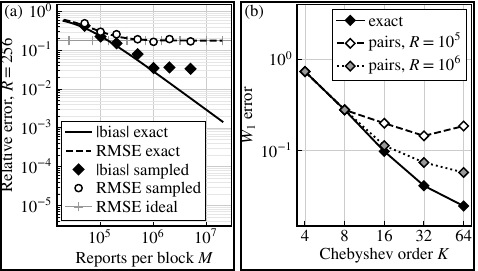}
\caption{Sensor-network graph: a weighted random geometric graph ($N=1024$, $4043$ edges, $C=17$, heat-diffused sensor map with $\tau=0.5$; Appendix~\ref{app:sensor}), logarithmic axes. (a) Absolute relative bias and relative RMSE of $\hat\cE$ versus reports per block $M$ for the sampled CB pipeline at $R=256$ ($40$ estimates per $M$; markers), the analytic tape-averaged values (lines) and the ideal-register RMSE (grey). (b) Wasserstein error versus Chebyshev order $K$ for exact moments and for pair-parity moments (Theorem~\ref{thm:spectrum}) from $R=10^5$ and $10^6$ ideal registers ($\bar\lambda=2\max_x\deg_w(x)$, $\epsilon_{\rm c}=0.01$; means of $20$ replicates).}\label{fig:sensor}
\end{figure}

The sensor-network graph, a weighted random geometric graph (Appendix~\ref{app:sensor}), has $N=1024$ sites, $4043$ edges with weights $0.14$--$1.0$, degrees $1$ to $17$, $C=17$ classes and $\cE_G=1.99$. Fig.~\ref{fig:sensor}(a) runs the sampled CB pipeline on it with $R=256$ and $M=5.0\times10^{4}$ to $5.0\times10^{6}$ reports per block ($\kappa_bv=0.86$ to $0.0086$). The analytic relative bias from \eqref{eq:meanrho} falls from $0.38$ to $0.0058$, following $1/M$ within $11\%$ for $M\ge2\times10^5$. The analytic relative RMSE falls from $0.43$ to $0.178$, compared with $0.177$ for ideal registers. Sampled estimates ($40$ per $M$) have relative bias $+0.43$ to $+0.033$ and relative RMSE $0.50$ to $0.170$. These are within $1.4$ standard errors of the analytic bias and $1.8$ of the analytic RMSE at every $M$ (Appendix~\ref{app:jackknife}).

Fig.~\ref{fig:sensor}(b) applies the pair-parity spectral estimator to the same map ($\lambda_N=8.35$) with the disclosed bound $\bar\lambda=2\max_x\deg_w(x)=14.8$ of Section~\ref{sec:kpm}, where the exact-moment error falls from $0.74$ at $K=4$ to $0.025$ at $K=64$. The $K$-hop support graph has $33{,}688$ pairs at $K=4$ and is nearly complete from $K=16$ on. Thus $C_K$ grows from $99$ to $1023$ (torus: $40$ to $542$ for $K\le16$). The count \eqref{eq:Rspec} at $\varepsilon=\delta=0.05$ grows from $4.2\times10^{5}$ to $6.6\times10^{6}$ registers. Pair-parity moments at $R=10^5$ give errors of $0.15$--$0.19$ for $K\ge32$. The error at $R=10^6$ reaches $0.058$ at $K=64$. The standard errors of these means are at most $\figSensorWassSE$.

\subsection{End-to-end gate-level quantum circuit simulations}\label{sec:gatelevel}
\begin{figure}[!t]
\centering\includegraphics{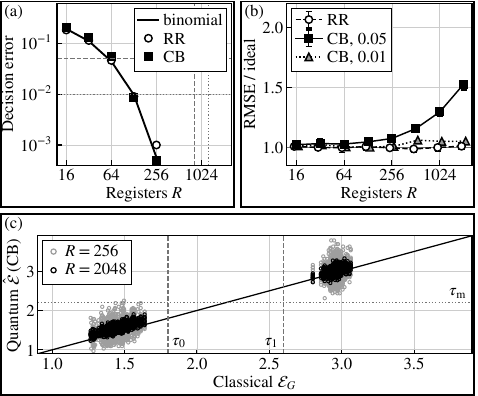}
\caption{Gate-level CB receiver on unit-weight tori ($b=4$, $C=4$, $\tau_0=1.8$, $\tau_1=2.6$), compared with the classical receiver on the same reports. (a) Decision error, the mean of the false-alarm and miss rates, versus $R$ at $N=2^{12}$, $\kappa_bv=0.05$, for CB and RR registers. It is the rate of disagreement between quantum and classical decisions. Markers show $1000$ trials per hypothesis and are omitted where no error occurred. The line shows the ideal-register binomial law. The targets are $\delta_s=0.05$ (dashed) and $0.01$ (dotted), with their counts \eqref{eq:Rbern} shown vertically. (b) RMSE of $\hat\cE$ about the classical $\cE_G$, divided by the ideal $\sqrt{(\E[Y_k^2]-\cE_G^2)/R}$: mean over $N=2^6$--$2^{16}$ (markers) and range (bars) for RR and CB registers at $\kappa_bv=0.05$, and CB registers at $\kappa_bv=0.01$, $N=2^{12}$. (c) Quantum estimate $\hat\cE$ of CB registers versus the classical $\cE_G$ in every trial of (a), at $R=256$ (grey) and $2048$ (black), with $\tau_0$, $\tau_1$ (dashed), $\tau_{\rm m}$ (dotted) and the diagonal.}\label{fig:gatelevel}
\end{figure}

This subsection runs Algorithm~\ref{alg:pipeline} in CB shared mode with gates, mid-circuit measurements and resets, one shot per register. It compares the decisions with those of a classical receiver storing all $Nb$ record bits from the same reports. Each register has $d$ address, $b=4$ label and one flag qubit. Epoch A applies one multi-controlled phase gate \eqref{eq:gate} per register and report inside the calls of Fig.~\ref{fig:circuit}. Epoch B tests membership in each edge of the register's class by multi-controlled NOT gates on the flag. A hit triggers the parity measurement. Qiskit Aer simulates these circuits for $N\le256$. A gate-by-gate simulator reproducing Qiskit's amplitudes to $2.4\times10^{-15}$ reaches $N=2^{19}$ ($24$ qubits per register). No outcome law, channel or operator matrix enters a run (Appendix~\ref{app:gatelevel} and Table~\ref{tab:gatesweep}). On unit-weight tori with $C=4$, smooth or rough maps ($\cE_G\le\tau_0=1.8$ or $\cE_G\ge\tau_1=2.6$) and $\kappa_bv=0.05$, \eqref{eq:Rbern} gives $R=819$ at $\delta_s=0.05$ and $1259$ at $0.01$. The classical receiver computes $\cE_G$ exactly. The quantum decision error is the rate of disagreement with the classical decision. Fig.~\ref{fig:gatelevel}(a) shows that the rate at $N=2^{12}$ follows the scores' binomial law for RR registers and stays close to it for CB registers. The rate at every size up to $N=2^{16}$ falls below $\delta_s=0.05$ by $R=128$ (CB at $N=64$: $256$) and below $0.01$ by $R=256$ (CB at $N=64$: $512$; Table~\ref{tab:gatesweep}). Maps at the thresholds give binomial-law counts of $252$ and $474$. The simulated pools have mean energies $1.43$--$1.57$ and $2.95$--$3.11$, which lie further apart. The distance $\delta_p$ for which Theorem~\ref{thm:prep}(ii) holds exceeds one at the simulated $M$ from $R=16$ on. These decisions show that its joint bound is conservative for tests with a gap. Fig.~\ref{fig:gatelevel}(b) shows the RMSE of $\hat\cE$ about the classical $\cE_G$. It stays within $5\%$ of the ideal $\sqrt{(\E[Y_k^2]-\cE_G^2)/R}$ for RR registers and within $6\%$ for CB registers at $\kappa_bv=0.01$. CB preparation bias at $\kappa_bv=0.05$ is $+7\%$ to $+8\%$ on smooth maps. It is visible from $R\approx512$ and raises the ratio to as much as $1.55$ at $R=2048$. Fig.~\ref{fig:gatelevel}(c) plots each quantum estimate against the classical $\cE_G$ of the same reports. Estimates scatter about the classical values, shifted by CB bias. They lie on the correct side of $\tau_{\rm m}$ in all but one of $2000$ trials at $R=256$ and in all at $R=2048$. No decision error occurred at $N=2^{19}$ with $20$ trials per hypothesis from $R=256$ on. The controller holds at most $828$ classical bits at $R=256$ and none across the cutoff. The simulated receiver's $R(d+b)+1$ qubits fall below the $Nb$ record bits only from $N=2^{10}$ at $R=256$. The separation is asymptotic.

\section{Conclusion}\label{sec:conclusion}
We studied smoothness and spectral estimation for fused one-bit sensor maps on a graph that is known only after the raw reports are erased. Our receiver combines coherent loading of the sub-bits, exact in RR and by QOS in CB, with nonlinear fusion and HM readout. We showed that, at fixed $b$, resolution, graph and kernel bounds and error targets, a register count independent of $N$, and $O(\log N)$ logical qubits and classical bits, suffice for smoothness tests, for estimates of the Dirichlet energy and kernel functionals and, with planner-supplied coefficients, for spectral distributions. In RR and CB, a classical record buffer before HM uses $O(Nb)$ bits, and every classical receiver needs $\Omega(\sqrt N)$ bits in the worst case; we prove this exponential memory separation. In simulations with $b=4$, the estimates of the broadcast pipeline match the analytic tape averages at $N=64$ and $1024$ and show no resolved bias growth up to $N=2^{20}$, and gate-level decision errors fall below $0.01$ by $R=512$. The separation concerns retained memory only: fault-tolerant hardware, gate rates and, in CB, report counts remain open.

\appendices
\section{Proof of Theorem~\ref{thm:prep}}\label{app:prep}
\emph{Single sub-bit reports (Section~\ref{sec:loading}).} Indexed-bit subsampling is a part of QOS \cite[Eqs.~(D91)--(D92)]{ZhaoQOS}. Algorithm~2 and Lemma~D.15 of \cite{ZhaoQOS} use complete records. The sparse-element XOR construction uses the entire observed bitstring \cite[Lemma~D.19, Eq.~(D145)]{ZhaoQOS}. Each one-bit device reports alone (RR and CB), so a report supplies only one sub-bit. The raw-report contract does not directly allow these complete-record updates (Section~\ref{sec:contracts}). The bounds in \cite[Thms.~D.12--D.13]{ZhaoQOS} imply that $2b$ controlled single-bit calls of indexed-bit subsampling suffice with about $4\pi^2Nb^3/\delta_p$ reports per register at trace distance $\delta_p$. This is about $8b/\kappa_b$ times the budget of Theorem~\ref{thm:prep}(i) at $R=1$.

\emph{Tape averaging and call error (Section~\ref{sec:tape}).} A call block applies $\widetilde{\mathbf X}_\omega$ for the arriving tape $\omega$. The tape is random and discarded. A tape-independent input $\boldsymbol\rho$ therefore leaves the block in the tape-averaged state, as in \cite[Eq.~(D54)]{ZhaoQOS}:
\begin{equation}
\mathcal C_M(\boldsymbol\rho)\triangleq\E_\omega[\widetilde{\mathbf X}_\omega\boldsymbol\rho\widetilde{\mathbf X}_\omega^\dagger]\label{eq:tapeavg}
\end{equation}
Thus $\mathcal C_M$ describes one call block. The ideal-prefix inputs $\boldsymbol\rho$ have uniform magnitude over $(x,\mathbf a)$ in the phase basis of the kicks \eqref{eq:kick}. The maximum trace distance of $\mathcal C_M$ from the ideal XOR on these inputs defines the call error as follows:
\begin{equation}
e_b\triangleq\max_{\boldsymbol\rho}D_{\rm tr}\big(\mathcal C_M(\boldsymbol\rho),\mathbf X_s\boldsymbol\rho\mathbf X_s^\dagger\big)\label{eq:callerror}
\end{equation}
It obeys \eqref{eq:singlecall}, where $\kappa_b\triangleq(1+\sqrt{1+1/b})/2$ and $v\triangleq\pi^2Nb^2/(4M)$. First, suppose every cell received exactly its mean count $M/(Nb)$ of reports. The kicks would give $\mathbf U_{\rm kick}$, hence the ideal XOR. Second, the random counts add to $\ket{x,\mathbf a}$ a raw phase error of zero mean and variance at most $v|\mathbf a|/b$. Each active label bit contributes at most $\theta^2M/(Nb)$. This equals $v/b$ at $\theta=\pi Nb/(2M)$, so $v$ is the phase-variance scale. Third, expand $e^{\mathrm i\epsilon}-1$ to $2$nd order in the phase error $\epsilon$ and average over the tape and the uniform labels of an ideal-prefix input. The trace-norm bounds are $\frac v2\sqrt{1+1/b}$ for the linear terms and $\frac v2$ for the quadratic term. Together these give $\kappa_bv/2$ in trace distance. The bound holds for the signed kicks and these inputs, not for arbitrary inputs. Fresh-block call-error bounds add along the ideal-prefix comparison.

A block of $M$ uniform reports has counts $\nu_{x,j}$ and signs $z_{x,j}=(-1)^{s_{x,j}}$. Its phase on $\ket{x,\mathbf a}$ is $\theta\sum_ja_jz_{x,j}\nu_{x,j}$. The gate $\mathbf S_L^\dagger$ and the label Hadamards turn it into a bitwise exclusive OR (XOR) at the mean counts \cite[Sec.~8.4]{BernsteinVazirani1997}. The raw phase error is $\epsilon^{\rm raw}_{x,\mathbf a}=\theta\sum_{j=1}^b a_jz_{x,j}(\nu_{x,j}-M/(Nb))$, with $\mathbf a\in\{0,1\}^b$ and Hamming weight $|\mathbf a|=\sum_{j=1}^ba_j$. Its mean is zero and $\E[(\epsilon^{\rm raw}_{x,\mathbf a})^2]\le v|\mathbf a|/b$ by the multinomial, or pairwise-uniform, covariance. Each call has uniform input magnitude in the $(x,\mathbf a)$ phase basis when all preceding calls are ideal. Let $\ket{\Psi_{\rm pre}}$ denote this normalized comparison input, and let $i=(x,\mathbf a)$ index its $m_{\rm ph}=N2^b$ phase-basis amplitudes. The expectation here averages the random tape $\omega$ unless indicated otherwise. The operators $\E_{\mathbf a}$ and $\E_i$ average uniform labels and basis indices. Such $\boldsymbol\rho=\ket{\Psi_{\rm pre}}\!\bra{\Psi_{\rm pre}}$ and $\mathbf E_{\rm ph}=\mathrm{diag}_i(e^{\mathrm i\epsilon^{\rm raw}_i}-1)$ give the state change:
\begin{align}
&\E[(\mathbf I+\mathbf E_{\rm ph})\boldsymbol\rho(\mathbf I+\mathbf E_{\rm ph})^\dagger]-\boldsymbol\rho\notag\\
&\quad=\E[\mathbf E_{\rm ph}]\boldsymbol\rho+\boldsymbol\rho\,\E[\mathbf E_{\rm ph}]^\dagger
       +\E[\mathbf E_{\rm ph}\boldsymbol\rho\mathbf E_{\rm ph}^\dagger].
\end{align}
The bounds $|\E[(\mathbf E_{\rm ph})_{ii}]|\le\E[(\epsilon^{\rm raw}_i)^2]/2$ and $\E_{\mathbf a}[|\mathbf a|^2]=b(b+1)/4$ bound the first two terms in trace norm by $2\|\E[\mathbf E_{\rm ph}]\ket{\Psi_{\rm pre}}\|\le\frac v2\sqrt{1+1/b}$. The last term is positive with trace at most $v/2$. Thus each call is within trace distance $\kappa_bv/2$ of the ideal XOR. The fusion phase is exact and the reset is a channel. Let hybrid $k$ apply the ideal calls $1,\dots,k$ and then the actual calls $k+1,\dots,2R$. Hybrids $k-1$ and $k$ differ only in call $k$, whose input is an ideal-prefix state. Later calls, with independent tapes, are channels and do not increase trace distance \cite[Thm.~9.2]{NielsenChuang2010}. The $2R$ call errors add, and $R\kappa_bv\le\delta_p$ gives (i). Each step uses only second moments of counts, so pairwise-independent uniform reports suffice. The common tapes in (ii) correlate the copies. Consider the $m_{\rm ph}=N2^b$ indices $i=(x,\mathbf a)$ and remove the global phase $\bar\epsilon^{\rm raw}=m_{\rm ph}^{-1}\sum_i\epsilon^{\rm raw}_i$. The centred errors $\epsilon^{\rm ctr}_i=\epsilon^{\rm raw}_i-\bar\epsilon^{\rm raw}$ have a positive semidefinite covariance $\boldsymbol\Sigma_{\rm ph}$, $(\boldsymbol\Sigma_{\rm ph})_{ij}=\E_\omega[\epsilon^{\rm ctr}_i\epsilon^{\rm ctr}_j]$. Thus $\boldsymbol\Sigma_{\rm ph}\mathbf 1=0$. Its mean diagonal $\bar v_{\rm ph}=m_{\rm ph}^{-1}\sum_i(\boldsymbol\Sigma_{\rm ph})_{ii}$ satisfies $\bar v_{\rm ph}\le v/2$ because $\E_{\mathbf a}[|\mathbf a|]=b/2$. Its maximum diagonal is at most $(\sqrt v+\sqrt{v/(4N)})^2=vL_N$ because $\operatorname{Var}[\bar\epsilon^{\rm raw}]\le v/(4N)$. Here $\mathbf1$ is the all-ones column vector. A uniform $R$-tuple of indices $(i_1,\dots,i_R)$ drawn independently with replacement gives the joint error $\eta_{\rm ph}=\sum_{r_{\rm cp}=1}^R\epsilon^{\rm ctr}_{i_{r_{\rm cp}}}$ with tape variance $V_{\eta}=\E_\omega[\eta_{\rm ph}^2]$. Averaging $V_{\eta}$ over the tuple gives $\E[V_{\eta}]=R\bar v_{\rm ph}$ and the second moment:
\begin{align}
\E[V_{\eta}^2]&=R\,\E_i[(\boldsymbol\Sigma_{\rm ph})_{ii}^2]+R(R-1)\bar v_{\rm ph}^2\notag\\
&\quad+2R(R-1)\tr\boldsymbol\Sigma_{\rm ph}^2/m_{\rm ph}^2,
\end{align}
where terms in which a copy appears once vanish by $\boldsymbol\Sigma_{\rm ph}\mathbf1=0$. The last term collects both pairings of two distinct copies. Thus $\E[V_{\eta}^2]\le v^2(RL_N/2+3R(R-1)/4)$, using $\tr \boldsymbol\Sigma_{\rm ph}^2\le(\tr \boldsymbol\Sigma_{\rm ph})^2$. Each call is then within $\frac12\E[V_{\eta}^2]^{1/2}+R\bar v_{\rm ph}/2\le\chi_{R,N}v$, and two calls give $2\chi_{R,N}v\le\delta_p$. Every count equals one in (iii). Each report gate computes a $d$-qubit equality flag with $d-1$ logical ANDs and uncomputes it by measurement. These $d-1$ ancillas and the $\max\{0,b-2\}$ of $\mathbf T_F$ (Appendix~\ref{app:runtime}) are used at different times. They give $n_{\rm aux}$ in (iv).

\emph{Proposition~\ref{prop:serial}.} Independent blocks give a product state of the registers. A register's two calls give its state $\boldsymbol\rho_A$ the distance $D_1=D_{\rm tr}(\boldsymbol\rho_A,\boldsymbol\psi_{\bz})\le2e_b\le\kappa_bv$. The identity $\operatorname{Tr}(\boldsymbol\rho_A-\boldsymbol\psi_{\bz})=0$ gives the bias bound for the mean score \eqref{eq:meanrho}: $|\operatorname{Tr}((\boldsymbol\rho_A-\boldsymbol\psi_{\bz})(\bL-\frac{\lambda_N}2\mathbf I))|\le\lambda_ND_1=\beta$. Equation~\eqref{eq:classeffect} gives $\E[Y_k^2]=4C\sum_ew_e^2\operatorname{Tr}(\boldsymbol\rho_A\ket{e,1}\!\bra{e,1})\le2C\wmax\E[Y_k]$ for every address state. The scores are independent. Their means are at most $\tau_0+\beta$ under $\mathcal H_0$ and at least $\tau_1-\beta$ under $\mathcal H_1$. The proof of Theorem~\ref{thm:dirichlet}(iii) with deviation $\gamma/2-\beta$ gives \eqref{eq:Rserial}.

\emph{Remark~\ref{rem:robust}.} RR kicks are diagonal in the address. The state before the reset is therefore $N^{-1/2}\sum_x\ket x\ket{\phi_x}$ with unit label vectors $\ket{\phi_x}$. Here $\ket{\phi_x}=z_x\ket{0^b}$ at every site with two complete and consistent rounds. An edge $e=\{u,v\}$ contributes $2w_e(1-\operatorname{Re}\langle\phi_u|\phi_v\rangle)/N\in[0,4w_e/N]$ to $\operatorname{Tr}(\boldsymbol\rho_A\bL)$ after the reset. It contributes $4w_eD_e/N$ if both ends are such sites. Only edges at the other $m$ sites change, each by at most $4w_e/N$ regardless of their signs. This proves the first bound. If both rounds give complete records with XOR $\boldsymbol\delta_x\ne0^b$ at one of the $m$ sites, $\ket{\phi_x}\propto\ket{\boldsymbol\delta_x}$ is orthogonal to $\ket{0^b}$. An edge from this site to a consistent site then contributes $2w_e/N$. Charging $2w_e/N$ per incident edge to each of the $m$ sites gives the second bound. CB may have cell probabilities $p_c$ and angles $\pi/(2Mp_c)$. The count of cell $c$ has mean $Mp_c$ and variance at most $Mp_c$. The mean phase is therefore $\pi/2$, and the cell's phase variance is at most $\pi^2/(4Mp_c)$. The proof above then holds with the phase-variance scale $v=\pi^2b/(4Mp_{\min})$.

\emph{(v).} Write $D(\phi)=\mathrm{diag}(e^{-\mathrm i\phi/2},e^{\mathrm i\phi/2})$, $z_j=(-1)^{s_{x,j}}$ and $\epsilon_j=\theta z_j(\nu_{x,j}-M/(Nb))$, with $\epsilon_j'$ for the unload tape. The centred kicks of cell $(x,j)$ act on label qubit $j$ as $D(\pi z_j/2+\epsilon_j)$. Since $\mathbf S^\dagger D(\pi z_j/2)=e^{-\mathrm i\pi z_j/4}\mathrm{diag}(1,z_j)$, a call acts as $e^{-\mathrm i\pi z_j/4}\mathbf X^{s_{x,j}}(\cos\frac{\epsilon_j}2\,\mathbf I-\mathrm i\sin\frac{\epsilon_j}2\,\mathbf X)$: the XOR, a phase and a bit flip. The label returns to $0^b$ when both calls flip the same set $f$ of bits. The gate $\mathbf T_F'$ cancels the phase $(-1)^{|\mathbf s_x|}$ of the two calls. Thus the amplitude of $\ket{x,0^b}$ is $N^{-1/2}e^{-\mathrm i\pi b/2}z_xq_x$. Use $s_j=\sin\frac{\epsilon_j}2$, $c_j=\cos\frac{\epsilon_j}2$, primes for the unload tape and $\sigma_x(f)=(-1)^{F(\mathbf s_x\oplus f)+F(\mathbf s_x)}$. Then:
\begin{equation}
q_x=\textstyle\sum_{f}\sigma_x(f)\prod_{j\in f}s_js_j'\prod_{j\notin f}c_jc_j'\label{eq:qx}
\end{equation}
The kept state is thus $\mathbf K\ket{\psi_{\bz}}$ with $\mathbf K=\mathrm{diag}(q_x)$. Its squared norm is $P=N^{-1}\sum_xq_x^2$, and $\boldsymbol\rho_{\rm keep}=\E[\mathbf K\boldsymbol\psi_{\bz}\mathbf K]/\E[P]$. Let $a_x=1-q_x$ and $S_x=\sum_j(\epsilon_j^2+\epsilon_j'^2)$. The inequalities $|s_js_j'|+|c_jc_j'|\le1$, $1-\prod_ku_k\le\sum_k(1-u_k)$ on $[-1,1]$ and $1-|\cos y|\le y^2/2$ imply $0\le a_x\le S_x/4$. Define $m_x=\E[a_x]$, their mean $m$ over $x$ and $d_x=m_x-m$. Expanding $\mathbf K=\mathbf I-\mathrm{diag}(a_x)$ gives $\E[P]=1-2m+N^{-1}\sum_x\E[a_x^2]$. The first-order term $-2m\boldsymbol\psi_{\bz}$ is common to all sites, giving:
\begin{equation}
D_{\rm tr}(\boldsymbol\rho_{\rm keep},\boldsymbol\psi_{\bz})\le\frac{N^{-1}\sum_x\E[a_x^2]+(N^{-1}\sum_xd_x^2)^{1/2}}{\E[P]}\label{eq:keepbound}
\end{equation}
A cell count is binomial with mean $\lambda=M/(Nb)$, variance at most $\lambda$ and fourth central moment at most $3\lambda^2+\lambda$. Two cells' squared deviations have covariance at most $2\lambda^2/(Nb)^2$. Set $u=v/b$. Here $\lambda>12$ for $v\le\frac15$. These bounds give $\E[\epsilon_j^2]\le u$, $\E[\epsilon_j^4]\le u^2(3+1/\lambda)$, $\E[S_x]\le2v$ and $\E[S_x^2]\le4v^2+6v^2/b$. Hence $\E[a_x^2]\le v^2(\frac14+\frac3{8b})$ and $\E[P]\ge1-2m\ge1-v$. The tapes are independent, so $m_x=1-\sum_f\sigma_x(f)g_f^2$ with $g_f=\E[\prod_{j\in f}s_j\prod_{j\notin f}c_j]$. The quantity $g_f^2$ depends on $|f|$ only, so $|d_x|\le2\sum_{f\ne\emptyset}g_f^2$. When $|f|=1$, $\E[\epsilon_1]=0$ gives $|g_f|\le\E[|\epsilon_1|\sum_{j>1}\epsilon_j^2]/16+\E|\epsilon_1|^3/48\le bu^{3/2}(3+1/\lambda)^{1/2}/16$. Also $\sum_{|f|\ge2}g_f^2\le\E[(\sum_j|s_js_j'|)^2]\le b^2u^2/16$. Therefore $|d_x|\le v^2/8+0.0241v^3$, and \eqref{eq:keepbound} is at most $v^2(\frac38+\frac3{8b}+0.0241v)/(1-v)\le v^2$. Fresh blocks give independent kept registers, so $R$ of them are within $Rv^2$ of ideal copies. Exact tape averages for $N=16$--$128$, $b=1$--$5$, $v=0.005$--$0.2$ and threshold, parity and OR fusion give $0.82$--$0.99$ of $v^2/(8b)$ and discard probabilities $0.85$--$1.00$ of $v/2$.

\section{Proofs of Lemma~\ref{lem:outcome} and Theorems~\ref{thm:dirichlet} and~\ref{thm:kernel}}\label{app:est}
\emph{Lemma~\ref{lem:outcome}.} The $\mathbf P_e$, $e\in M_c$, have disjoint supports. The state $\ket{\psi_{\bz}}$ gives $\Pr(\mathbf P_e)=2/N$. The post-hit state $(z_u\ket u+z_v\ket v)/\sqrt2$ then gives $\Pr(r\mid e)=(z_u+(-1)^rz_v)^2/4$.

\emph{Theorem~\ref{thm:dirichlet}.} Each edge is hit with probability $2/(CN)$, so $\E[Y_k]=\sum_e\frac2{CN}2Cw_eD_e=\cE_G$ and $\E[Y_k^2]=\sum_e\frac2{CN}(2Cw_e)^2D_e\le2C\wmax\cE_G$. Bernstein's inequality \cite[Cor.~2.11]{Boucheron2013} applies to the centred independent and identically distributed (IID) scores. The hypothesis $\mathcal H_0$ uses deviation $\gamma/2$ and $\cE_G\le\tau_0$. Let $\cE$ vary. The hypothesis $\mathcal H_1$ uses $\cE_G-\tau_{\rm m}$. The ratio $(\cE-\tau_{\rm m})^2/[\cE+(\cE-\tau_{\rm m})/3]$ increases for $\cE\ge\tau_1>\tau_{\rm m}\ge0$, so its minimum is at $\tau_1$. The case $\tau_{\rm m}\le0$ is trivial. This gives \eqref{eq:Rbern}. Two tails give (iv), and Eq.~\eqref{eq:transfer} transfers the full event to the prepared state.

\emph{Theorem~\ref{thm:kernel}.} Each support edge is hit with probability $2/(C_{\bH}N)$ and $\E[(-1)^r\mid e]=z_uz_v$. Thus $\E[Z_k]=(2/N)\sum_eH_ez_uz_v$, and Lemma~\ref{lem:kernel} gives unbiasedness. Also $\E[Z_k^2]=(C_{\bH}/N)\|\bH_{\rm off}\|_F^2\le C_{\bH}\overline{h^2}$. Bernstein's inequality with $|Z_k-\E[Z_k]|\le2C_{\bH}h_{\rm off}$ concludes.

\emph{Class allocation (Section~\ref{sec:estimators}).} Fix positive counts $R_c$. Define $\bar Y_c=R_c^{-1}\sum_{k:c_k=c}Y_k$, $\cE_c=(4/N)\sum_{e\in M_c}w_eD_e$ and $\sigma_c^2=(8C^2/N)\sum_{e\in M_c}w_e^2D_e-C^2\cE_c^2$. The registers of class $c$ are IID with mean $C\cE_c$ and variance $\sigma_c^2$. Thus the class-mean estimator $\hat\cE=C^{-1}\sum_c\bar Y_c$ is unbiased with $\operatorname{Var}[\hat\cE]=\sum_c\sigma_c^2/(C^2R_c)$. Equal allocation $R_c=R/C$ gives $R\operatorname{Var}[\hat\cE]=C^{-1}\sum_c\sigma_c^2$. This scaled variance lies below the random-class value by $C^2\operatorname{Var}_c[\cE_c]$, by the law of total variance \cite[Sec.~4.3]{bertsekas2008introduction} applied over the uniform class choice.

\section{Proofs of Theorems~\ref{thm:classical} and~\ref{thm:ring}, caches and early graphs}\label{app:classical}
\emph{Hard family (Section~\ref{sec:classical}).} This family embeds the partial-matching problem $\alpha$PM of \cite[Sec.~2]{GKKRW} at $\alpha=\frac14$. The common hidden answer $\beta$ is not given to the planner. Here $B\in\{\mathsf L,\mathsf R\}$ labels the left or right bank, $i=1,\ldots,n$ a source coordinate and $\varsigma\in\{0,1\}$ its copy bit. Here $[n]$ means $\{1,\ldots,n\}$.
Let $n\ge4$ be a power of two, $\beta\in\{0,1\}$, $\mathbf x\in\{0,1\}^n$, and $\mathcal P$ a set of $n/4$ disjoint pairs with bits $\zeta_{uv}=x_u\oplus x_v\oplus\beta$. The $N=4n$ sites are $(B,i,\varsigma)\in\{\mathsf L,\mathsf R\}\times[n]\times\{0,1\}$ with statuses $x_i\oplus\varsigma$. Define unit-weight classes; $\boldsymbol\zeta$ collects Bob\textquotesingle s bits $\zeta_{uv}$. $M_1$ contains $(\mathsf L,u,\varsigma)$--$(\mathsf R,v,\varsigma\oplus \zeta_{uv})$ and $(\mathsf L,v,\varsigma)$--$(\mathsf R,u,\varsigma\oplus \zeta_{uv})$ for $\{u,v\}\in\mathcal P$, and $(\mathsf L,i,\varsigma)$--$(\mathsf R,i,\varsigma)$ for unpaired $i$. $M_2$ contains $(\mathsf L,i,\varsigma)$--$(\mathsf R,i,\varsigma\oplus1)$. $M_3$, $M_4$ contain $(B_i,i,0)$--$(B_i,i+1,0)$ and $(B_i,i,0)$--$(B_i,i+1,1)$, respectively (fan $i$) for $i\in[n-1]$. Here $B_i=\mathsf L$ for odd and $\mathsf R$ for even $i$. They properly colour a simple connected late graph $G_n(\mathcal P,\boldsymbol\zeta)$ of maximum degree $4$, $|E|=6n-2$ edges and $\cE_G=3+\beta-1/n$.

\emph{Properties.} $M_1$ and $M_2$ join the banks. Cross edges join different indices; identity and $M_2$ edges join equal indices with equal and opposite $\varsigma$. Fan edges join same-bank sites of consecutive indices, so the graph is simple. $M_1$ and $M_2$ are perfect matchings. Fan centres $(B_i,i,0)$ are never leaves of fan $i-1$, whose leaves lie in bank $B_{i-1}\ne B_i$. Thus $M_3$ and $M_4$ are matchings and the maximum degree is four. $M_2$ splits each cluster $i$ (the four sites $(B,i,\varsigma)$) into two components. Fan $i$ joins both components of cluster $i+1$ to $(B_i,i,0)$. An $M_1$ edge attaches the remaining component of cluster $1$, so the graph is connected. $M_1$ contributes $n\beta$ disagreements, $M_2$ contributes $2n$, and each fan contributes one. The edge count is $|E|=2n+2n+2(n-1)$.

\emph{Proof of Theorem~\ref{thm:classical}.} Alice holds $\mathbf x$ and Bob $(\mathcal P,\boldsymbol\zeta)$ under the promise of $\alpha$PM. Alice places records $\mathbf r_0,\mathbf r_1$ of fused values $0,1$ at the sites $(B,i,\varsigma)$ according to $x_i\oplus\varsigma$. She generates epoch A from these records alone, runs the receiver to the cutoff, and sends its boundary memory. Bob streams the colour-labelled edges of $M_1$, of energy $\beta$, for (a), or of $G_n(\mathcal P,\boldsymbol\zeta)$, of energy $3+\beta-1/n$, for (b). He reads $\beta$ from the decision. His stream and its colours depend on $(\mathcal P,\boldsymbol\zeta)$ only. Public parameters depend on $n$ only. Alice draws epoch-A randomness herself under either contract. This problem on $n$ bits needs $\Omega(\sqrt n)$ bits of one-way communication at error $\frac13$ \cite[Thm.~2]{GKKRW}, where $N=4n$.

\emph{Site cache (Section~\ref{sec:classical}).} Fix a graph and an immutable raw table, and choose an integer $1\le s_{\rm c}\le N=2^d$. Let $\mathcal S$ contain each site with probability $p_{\rm c}=s_{\rm c}/N$ through a four-wise independent hash with $O(d)$ seed bits, independent of the input and report process. Identify sites with $\mathrm{GF}(2^d)$ and draw four independent uniform field elements $a_0,\ldots,a_3$. Include $x$ when $a_0+a_1x+a_2x^2+a_3x^3$ lies in a fixed $s_{\rm c}$-element subset. The Vandermonde determinant gives independent uniform values at any four distinct sites. This gives exact inclusion with $4d$ seed bits. The cache keeps $(x,\mathbf s_x)$ and $b$ seen-flags for $x\in\mathcal S$ during a complete A-scan. It forms the uncapped complete-data estimate $\hat\cE_{\rm cl}=4(Np_{\rm c}^2)^{-1}\sum_{e\subseteq \mathcal S}w_eD_e$ over epoch-B edges. Conditioned on the table and graph, the indicators $I_x=\mathbf 1\{x\in\mathcal S\}$ give $X_e=w_eD_eI_uI_v$ with $\E[X_e]=p_{\rm c}^2w_eD_e$, so $\E[\hat\cE_{\rm cl}]=\cE_G$. Four-wise independence makes disjoint edges uncorrelated. Also $\operatorname{Var}[X_e]\le p_{\rm c}^2w_e^2D_e$, and adjacent-edge covariance is at most $p_{\rm c}^3w_ew_{e'}D_eD_{e'}$. Each edge has at most $2(\Delta-1)$ neighbours, and $\sum_ew_eD_e=N\cE_G/4$. Combining these bounds gives the variance bound:
\begin{equation}
\operatorname{Var}[\hat\cE_{\rm cl}]\le\big(4\wmax\cE_GN/s_{\rm c}^2\big)\big(1+2(\Delta-1)s_{\rm c}/N\big)\label{eq:cachevar}
\end{equation}
The estimate is exact for edgeless graphs. Chebyshev's inequality \cite[Sec.~2.1]{Boucheron2013} uses deviation at least $\gamma/2$ under the lower hypothesis and the actual deviation $\cE_G-\tau_{\rm m}$ under the upper one. The ratio $\cE_G/(\cE_G-\tau_{\rm m})^2$ decreases on $\cE_G\ge\tau_1$ if $\tau_{\rm m}\ge0$; the case $\tau_{\rm m}\le0$ is trivial. The midpoint rule, alarming iff $\hat\cE_{\rm cl}\ge\tau_{\rm m}$, therefore has per-hypothesis error at most $\delta_s$ if $\Delta\ge1$ and $s_{\rm c}\ge4\gamma^{-1}[\wmax\tau_1N(1+2(\Delta-1)s_{\rm c}/N)/\delta_s]^{1/2}$. Thus $s_{\rm c}=O(\sqrt N)$ at fixed $C$, $\wmax$, $\gamma$ and $\delta_s$. The expected record payload is $s_{\rm c}(d+2b)$ bits ($s_{\rm c}(d+b)$ in RR, where seen-flags are not needed). Seed, control and arithmetic are additional. A capacity of $h_{\rm c}=\min\{N,2s_{\rm c}\}$ sites overflows with probability at most $\delta_{\rm cap}=\min\{1,3/s_{\rm c}^2+1/s_{\rm c}^3\}$, zero if $h_{\rm c}=N$. The selected-site count $n_{\rm cache}$ has $\E[(n_{\rm cache}-s_{\rm c})^4]\le s_{\rm c}+3s_{\rm c}^2$. The fourth-moment Markov inequality \cite[Sec.~2.1]{Boucheron2013} then gives the overflow bound. Suppose epoch A delivers $T_A$ IID reports on uniformly random (site, bit) cells, independent of the hash. A union bound over at most $h_{\rm c}b$ cached cells gives the missing-record allowance $\delta_{\rm miss}=\min\{1,2s_{\rm c}b\,e^{-T_A/(Nb)}\}$. The capped decision error is at most $\delta_s+\delta_{\rm cap}+\delta_{\rm miss}$; complete scans set $\delta_{\rm miss}=0$. A sufficient size above $N$ means full caching and exact energy. The sufficient size in Fig.~\ref{fig:memory}(a) is plotted in bits and uncapped; its site count exceeds $N$ at both $N=2^{10}$ and $N=2^{12}$.

\emph{Early graphs (Section~\ref{sec:classical}).} Suppose $G$ is disclosed before epoch A and $W=\sum_ew_e>0$. Draw $k_{\rm s}\ge1$ edges independently with replacement, with probabilities $w_e/W$. A once-only early graph stream uses $k_{\rm s}$ independent one-edge reservoirs. Each replaces its edge at a positive-weight edge $e_j$ with probability $w_{e_j}/W_j^{\rm pre}$, $W_j^{\rm pre}=\sum_{i\le j}w_{e_i}$. Induction gives the marginal $w_e/W$, and independent coins give the product law. Complete acquisition of the at most $2k_{\rm s}$ endpoint records gives scores $(4W/N)D_{e_i}$ with mean $\cE_G$ and range $[0,4W/N]$. The average of the $k_{\rm s}$ scores is unbiased. Hoeffding's inequality \cite[Thm.~2]{Hoeffding1963} on both tails places it within $\varepsilon$ of $\cE_G$ with probability at least $1-\delta_s$ for $k_{\rm s}\ge8(W/N)^2\varepsilon^{-2}\ln(2/\delta_s)$. A public bound such as $W/N\le C\wmax/2$ keeps $k_{\rm s}$ from growing with $N$. The record payload is $O(k_{\rm s}(d+b))$ bits; weights, arithmetic, counters and fusion control are additional. The case $W=0$ gives zero exactly. Take a fresh IID suffix of $M_{\rm suf}$ uniform raw-cell reports, independent of the history after fixing the samples. It misses a given endpoint cell with probability at most $e^{-M_{\rm suf}/(Nb)}$, adding failure probability at most $\min\{1,2k_{\rm s}b\,e^{-M_{\rm suf}/(Nb)}\}$.

\emph{Patch cache (Section~\ref{sec:classical}).} Consider a known $n_{\rm tor}\times n_{\rm tor}$ torus with $N=n_{\rm tor}^2=2^d$, $n_{\rm tor}\ge4$, promised as the unit-weight nearest-neighbour late graph. Choose $k_{\rm s}\ge1$ independent uniform origins of $\ell_{\rm p}\times\ell_{\rm p}$ blocks, $2\le\ell_{\rm p}<n_{\rm tor}$, before acquisition and independently of the table. Keep their records. An edge's first endpoint has $\ell_{\rm p}-1$ internal column and $\ell_{\rm p}$ row positions in a translated block. A uniform origin includes each horizontal or vertical edge with probability $\ell_{\rm p}(\ell_{\rm p}-1)/N$. The average over the blocks of $c_{\ell_{\rm p}}\mathrm{dis}_{\ell_{\rm p}}$, with $c_{\ell_{\rm p}}=4/(\ell_{\rm p}(\ell_{\rm p}-1))$ and $\mathrm{dis}_{\ell_{\rm p}}$ the disagreeing internal edges, is unbiased for $\cE_G$. Independent origins divide the single-block variance by $k_{\rm s}$, even when blocks overlap. A block has $2\ell_{\rm p}(\ell_{\rm p}-1)$ internal edges, so each score lies in $[0,8]$. Hoeffding's inequality \cite[Thm.~2]{Hoeffding1963} on both tails gives error at most $\varepsilon$ with failure at most $\delta_s$ for $k_{\rm s}\ge\lceil32\varepsilon^{-2}\ln(2/\delta_s)\rceil$, independent of $N$. This uses $k_{\rm s}\ell_{\rm p}^2(d+2b)$ record bits plus control. A complete scan supplies the records, or $M_{\rm suf}$ IID uniform reports after the origins are fixed add the allowance $\min\{1,b\min(N,k_{\rm s}\ell_{\rm p}^2)e^{-M_{\rm suf}/(Nb)}\}$.

\emph{Proof of Theorem~\ref{thm:ring}.} The problem $\alpha$PM at $\alpha=1/4$ \cite[Thm.~2]{GKKRW} gives Alice $\mathbf x\in\{0,1\}^n$ and Bob $n/4$ disjoint pairs with bits $\zeta_e$ such that $x_u\oplus x_v\oplus\zeta_e=\beta$. Its bounded-error one-way communication needs $\Omega(\sqrt n)$ bits. Take $4n$ old sites $(\mathsf L,i,\varsigma)$, $(\mathsf R,i,\varsigma)$ with statuses $x_i\oplus\varsigma$. Each pair connects $(\mathsf L,u,\varsigma)$ to $(\mathsf R,v,\varsigma\oplus\zeta_e)$ and $(\mathsf L,v,\varsigma)$ to $(\mathsf R,u,\varsigma\oplus\zeta_e)$ for both $\varsigma\in\{0,1\}$. Give unmatched indices identical-copy rungs. These edges form a bijection between the two $2n$-site banks. The $n$ matched-coordinate rungs have disagreement $\beta$; the other $n$ agree. Replace every old status $t$ by the four-site motif $\Phi(t)=(0,t,1,t)$ in a public block order independent of Bob. The three internal edges and the edge to the next block's initial zero give $t+(1-t)+(1-t)+t=2$ disagreements, including the cyclic closure. Thus $N=16n$, $D_{\rm cyc}=N/2$ and each cycle class has $N/4$ disagreements. Each old rung, lifted slot by slot, gives two agreeing slots and two that copy its disagreement. The $N/2$ lifted rungs therefore contain $2n\beta$ disagreements, $q_\varpi=\beta/4$ and $\cE_\varpi=(1+\beta)/2$. Place records $\mathbf r_0,\mathbf r_1$ of fused values $0,1$ according to the lifted statuses. The reduction of Theorem~\ref{thm:classical} now has Bob generate only the lifted rungs and their class from his pairs and $\boldsymbol\zeta$. It decides $\beta$ with error at most $\frac13$, so the boundary memory is $\Omega(\sqrt n)=\Omega(\sqrt N)$ bits. The quantum count uses $C=3$, $\wmax=1$, $\tau_1=1$ and $\gamma=1/2$ in \eqref{eq:Rbern}. Preparation and other errors take separate allowances, and \eqref{eq:certificate} does not cover this family.

\section{Allocation certificate}\label{app:cert}
Bernstein's inequality \cite[Cor.~2.11]{Boucheron2013} gives, for independent ideal copies and $\varepsilon>0$, $\Pr\{|\hat\cE-\cE_G|\ge\varepsilon\}\le2\exp[-\varepsilon^2/(2V_{\rm alloc}+2B_{\rm inc}\varepsilon/3)]$ with $V_{\rm alloc}=\operatorname{Var}[\hat\cE]$ and $B_{\rm inc}=\max_c2\wmax/R_c$. Each admissible instance of the hard family of Section~\ref{sec:classical} has class odd-hit probabilities determined by the hidden bit $\beta\in\{0,1\}$, the input size $n=N/4$, and the number $j_{\rm fan}$ of bit transitions. The class probabilities of an odd hit ($r_k=1$) are $(\beta/2,1,j_{\rm fan}/(2n),(n-1-j_{\rm fan})/(2n))$. The midpoint rule with $R_c=28$ alarms iff the total odd count reaches $49$ when $N\ge64$. Take independent counts $\mathsf B_3\sim\mathrm{Bin}(28,j_{\rm fan}/(2n))$, $\mathsf B_4\sim\mathrm{Bin}(28,(n-1-j_{\rm fan})/(2n))$, and $\mathsf B_1\sim\mathrm{Bin}(28,\frac12)$. The false-alarm (FA) and miss probabilities are:
\begin{equation}
\thickmuskip=2mu plus 2mu\medmuskip=1.5mu plus 1mu P_{\rm FA}=\Pr(\mathsf B_3+\mathsf B_4\ge21);P_{\rm miss}=\Pr(\mathsf B_1+\mathsf B_3+\mathsf B_4\le20)\label{eq:tails}
\end{equation}
For every admissible instance with $N\ge128$:
\begin{equation}
P_{\rm FA}<0.02674,\qquad P_{\rm miss}<0.04282\label{eq:certificate}
\end{equation}
and at $N=64$, $R=116$ gives $P_{\rm FA}<0.0195$ and $P_{\rm miss}<0.0327$. The certificate covers ideal copies of this family only.

Class $M_1$ has $n\beta$ disagreeing gadget edges, $M_2$ has $2n$, and each of the $n-1$ fans has one disagreement. It lies in $M_3$ at a transition of adjacent input bits and in $M_4$ otherwise. Let $j_{\rm fan}=\sum_{i=1}^{n-1}\mathbf 1\{x_i\ne x_{i+1}\}$ count adjacent-bit transitions. This gives the odd probabilities above with $p_3=s_{{\rm fan},n}t_{\rm fan}$ and $p_4=s_{{\rm fan},n}(1-t_{\rm fan})$, where $s_{{\rm fan},n}=(n-1)/(2n)\in[\frac{31}{64},\frac12)$ for $n\ge32$ and $t_{\rm fan}=j_{\rm fan}/(n-1)$. Consider a fan scale $s_{\rm fan}\in[31/64,1/2]$ and $t_{\rm fan}\in[0,1]$. Give $\mathsf B_3$ and $\mathsf B_4$ the probabilities $s_{\rm fan}t_{\rm fan}$ and $s_{\rm fan}(1-t_{\rm fan})$. Define the false-alarm (FA) and missed-detection (MD) tails $f^{\rm FA}_{s_{\rm fan}}(t_{\rm fan})=\Pr(\mathsf B_3+\mathsf B_4\ge21)$ and $f^{\rm MD}_{s_{\rm fan}}(t_{\rm fan})=\Pr(\mathsf B_1+\mathsf B_3+\mathsf B_4\le20)$. These are the fixed 112-register rule\textquotesingle s two tails in \eqref{eq:tails}. Write $\upsilon\in\{\mathrm{FA},\mathrm{MD}\}$ for either tail. Couple the Bernoulli trials through common uniforms. The false-alarm event is increasing and the miss event decreasing in $s_{\rm fan}$ at fixed $t_{\rm fan}$. It suffices to bound them at $s_{\rm fan}=\frac12$ and $s_{\rm fan}=\frac{31}{64}$ for all $t_{\rm fan}\in[0,1]$. Each tail obeys this bound for $t_{\rm fan},t_{\rm fan}^{\prime}\in[0,1]$: $|f^{\upsilon}_{s_{\rm fan}}(t_{\rm fan})-f^{\upsilon}_{s_{\rm fan}}(t_{\rm fan}^{\prime})|\le56s_{\rm fan}|t_{\rm fan}-t_{\rm fan}^{\prime}|$. The grid $t_{{\rm fan},q_{\rm grid}}=q_{\rm grid}/J_{\rm grid}$ with $J_{\rm grid}=16{,}384$ therefore gives the following:
\begin{align}
&\sup_{t_{\rm fan}\in[0,1]} f^{\upsilon}_{s_{\rm fan}}(t_{\rm fan})\notag\\
&\quad\le\max_{0\le q_{\rm grid}\le J_{\rm grid}}
 f^{\upsilon}_{s_{\rm fan}}(q_{\rm grid}/J_{\rm grid})+28s_{\rm fan}/J_{\rm grid},
\end{align}
where exact integer arithmetic on the grid values gives false-alarm and miss maxima below $0.0258775$ and $0.0419899$. The allowances $7/8192$ and $217/262144$ then give \eqref{eq:certificate}.

The unchanged hard-family decision rule, \eqref{eq:certificate} and the joint-state event bound \eqref{eq:transfer} give error below $0.04282+\delta_p+\delta_{\rm phys}$ when physical implementation changes each event probability by at most $\delta_{\rm phys}$. A $5\%$ target leaves $0.00718$. Reserving $\delta_p=0.005$ and $\delta_{\rm phys}=0.002$ gives a conservative total below $0.04982$. These are assigned allowances, not measured errors. Take $b=4$ and $R=112$. The serial width $Q_{\rm ser}$ of \eqref{eq:widths} grows from $1{,}133$ to $4{,}523$ logical qubits as $N$ grows from $2^{10}$ to $2^{40}$. The RR scenario's shared width $Q_{\rm sh}$ grows from $1{,}577$ to $4{,}967$. The fitted \emph{expected} cache payload in Fig.~\ref{fig:memory}(a) grows from $8.3\times10^3$ to $7.3\times10^8$ bits over the same range. It is neither a capped memory guarantee nor a certified error bound.

\section{Proofs of Theorem~\ref{thm:spectrum} and Corollary~\ref{cor:specclassical}}\label{app:spectrum}
\emph{Class counts on growing graphs (Section~\ref{sec:joint}).} Lemma~\ref{lem:colour} makes lattice budgets polynomial in $K$. Let the maximum degree be $\Delta\ge2$, and write $\Delta_{(K)}$ for the maximum degree of the $K$-hop support graph. An available palette satisfies $C_K\le\Delta_{(K)}+1\le\Delta\sum_{j=0}^{K-1}(\Delta-1)^j+1$. The sufficient budget can be much larger on rapidly growing graphs.

\emph{Jackson coefficients.} The coefficients $g_j$ of \eqref{eq:kpmcdf} are the Jackson coefficients \cite[Eq.~(71)]{Weisse2006}. Translate its number of moments to $K+1$ and set $\alpha_{\rm J}=\pi/(K+2)$:
\begin{equation}
g_j=\frac{(K+2-j)\cos(j\alpha_{\rm J})+\sin(j\alpha_{\rm J})\,\cot \alpha_{\rm J}}{K+2}\in[0,1],
\label{eq:jackson}
\end{equation}
for $0\le j\le K$. The integration uses $\int_{-1}^{\cos\varphi}T_j(t)\,dt/(\pi\sqrt{1-t^2})=-\sin(j\varphi)/(j\pi)$ for $j\ge1$. The Jackson kernel is nonnegative and normalized. Each line has standard deviation at most $\sin \alpha_{\rm J}\le\pi/(K+2)$ in $y$ \cite[Eqs.~(73)--(75)]{Weisse2006}. This is not a support radius. The support $[\bar\lambda/2-a_\lambda,\bar\lambda/2+a_\lambda]$ of $\bar F_K$ is slightly larger than $[0,\bar\lambda]$ for small $\epsilon_{\rm c}$.

\emph{(i)} Each $T_j(\tilde\bL)$ is a degree-$j$ polynomial in $\bL$. Its off-diagonal support therefore lies in the $K$-hop graph for $0\le j\le K$ (Lemma~\ref{lem:kernel}). The $C_K$-class colouring serves all $j$ simultaneously. Equation~\eqref{eq:jointest} is the estimator of Theorem~\ref{thm:kernel} with $\bH=T_j(\tilde\bL)$, hence unbiased. The spectrum of $\tilde\bL$ lies in $(-1,1)$ and $|T_j|\le1$ there. Thus $\overline{h^2}=N^{-1}\sum_kT_j(y_k)^2\le1$ and $h_{\rm off}\le\|T_j(\tilde\bL)\|\le1$. Theorem~\ref{thm:kernel} at confidence $\delta/K$ and a union bound over $1\le j\le K$ give (i) under \eqref{eq:Rspec}.

\emph{(ii)} Equation~\eqref{eq:kpmcdf} gives $F_K^{\rm raw}-\bar F_K=-\frac2\pi\sum_{j=1}^Kg_j(\hat\mu_j-\mu_j)\sin(j\varphi)/j$ before processing. Here $0\le g_j\le1$ because $g_j=\sum_{\nu=0}^{K-j} b_\nu^{\rm J} b_{\nu+j}^{\rm J}$ as in \cite[Eqs.~(57), (66)--(67)]{Weisse2006}. The source uses $N=K+1$, $k=1$. Its $a_\nu$ correspond to our $b_\nu^{\rm J}$, with $b_\nu^{\rm J}=\sqrt{2/(K+2)}\sin[\pi(\nu+1)/(K+2)]$ for $0\le\nu\le K$. They are positive and $g_0=1$.

\emph{(iii)} The kernel polynomial method (KPM) measure with exact moments is $\sum_kp_k\,J_K(\cdot\,;y_k)$. Its kernel, for $-1<y<1$, is $J_K(y;y_k)=[1+2\sum_{j=1}^K g_jT_j(y)T_j(y_k)]/(\pi\sqrt{1-y^2})$, as in \cite[Eqs.~(20), (72)]{Weisse2006}. The kernel $J_K$ vanishes outside $(-1,1)$. The broadened line $J_K(\cdot\,;y_k)$ is a probability density with mean $g_1y_k$. Its variance $\sigma_{\rm J}^2(y_k)=\int_{-1}^1(y-g_1y_k)^2J_K(y;y_k)\,dy$ obeys $\sigma_{\rm J}^2(y_k)\le\sin^2\frac{\pi}{K+2}$, and $g_1=\cos\frac{\pi}{K+2}$ \cite[Eqs.~(68), (71), (73)--(75)]{Weisse2006}. The source\textquotesingle s coordinate $x$ is our $y$, its line centre $a$ is our $y_k$ and its $N$ is $K+1$. Coupling each exact line with its broadened density yields the Wasserstein bound in $y$:
\begin{align}
W_1&\le\sum_kp_k\big(\sigma_{\rm J}^2(y_k)+(1-g_1)^2y_k^2\big)^{1/2}\notag\\
&\le\big(\sin^2\alpha_{\rm J}+(1-\cos\alpha_{\rm J})^2\big)^{1/2}=2\sin\frac{\pi}{2K+4}.
\end{align}
Moment errors before processing add the following:
\begin{align}
&\int_{-1}^1|F_K^{\rm raw}-\bar F_K|\,dy\notag\\
&\le\frac2\pi\sum_{j=1}^K\frac{g_j|\hat\mu_j-\mu_j|}j
\int_0^\pi|\sin j\varphi|\sin\varphi\,d\varphi\le\frac4\pi H_K\varepsilon,
\end{align}
and $d\lambda=a_\lambda\,dy$ converts to graph-frequency units. Clipping to $[0,1]$ cannot increase the pointwise distance to $\bar F_K\in[0,1]$. Increasing rearrangement cannot increase the uniform or $L^1$ distance to the nondecreasing $\bar F_K$, by \cite[Prop.~1(1)]{chernozhukov2008improving} with $p=\infty,1$. Its coordinate is $x=(y+1)/2$, its target is $f_0(x)=\bar F_K(2x-1)$, and its initial estimate $\hat f(x)$ is the clipped raw curve at $2x-1$. The $L^1$ change of variable multiplies both sides by two. Right-continuity and the continuous target extend the essential-supremum bound to the uniform bound, with the prescribed endpoint extensions. Thus (ii) and (iii) hold for the processed $\hat F_K$, a distribution function. The width is \eqref{eq:widths}, and \eqref{eq:transfer} transfers (i)--(iii) to prepared registers.

\emph{Corollary~\ref{cor:specclassical}.} Two laws with finite means and distribution functions $\mathcal F_1,\mathcal F_2$ have means differing by $\int_{\mathbb R}(\mathcal F_2-\mathcal F_1)\,d\lambda$, at most $W_1$ in size by \eqref{eq:wasserstein}. A law at finite $W_1$ distance from $F_{\bz}$ has a finite mean. The mean of $F_{\bz}$ is $\cE_G$ by Lemma~\ref{lem:dirichlet}. An estimate within $W_1$ distance $\varepsilon_{\rm W}<\frac12$ therefore gives $\cE_G$ within $\varepsilon_{\rm W}<\gamma/2$. It falls on the correct side of the midpoint and decides the hard family, so Theorem~\ref{thm:classical}(b) applies.

\section{Proofs of Lemmas~\ref{lem:dirichlet} and \ref{lem:colour}}\label{app:lemmas}
\emph{Lemma~\ref{lem:dirichlet}.} The edge form is $\bz^{\mathsf T}\bL\bz=\sum_ew_e(z_u-z_v)^2$, with $(z_u-z_v)^2=4D_e$. Orthonormality gives the spectral form and $\sum_kp_k=1$. If $\mathbf f$ is real, $(f_u-f_v)^2\le2(f_u^2+f_v^2)$ yields $\mathbf f^{\mathsf T}\bL \mathbf f\le2(\max_x\deg_w(x))\|\mathbf f\|^2$. This bounds $\lambda_N$.

\emph{Lemma~\ref{lem:colour}.} Alternate two colours on an even cycle. Use two along the rows and two along the columns on a torus with even side lengths. Here $K$-hop kernels use only nonzero offsets, denoted $o$ in one dimension and $\mathbf o$ in two, identifying residues modulo sign and period. Each offset generates translation orbits of even power-of-two length. Alternate two colours if the length is at least four, or use one for order two. Distinct offset pairs receive disjoint palettes. Before modular identification there are $K$ sign-pairs with $1\le|o|\le K$ in one dimension and $\sum_{j=1}^K2j=K(K+1)$ with $1\le\|\mathbf o\|_1\le K$ in two dimensions. Identification can only reduce these counts. The case $K=0$ has empty off-diagonal support.

\section{Gate counts and factory rates}\label{app:runtime}
\emph{Gate and operation counts (Section~\ref{sec:complexity}).} Within the $n_{\rm aux}$ ancillas of \eqref{eq:widths}, a kick follows \eqref{eq:gate}. It computes the flag of address $x$ with $d-1$ logical ANDs, applies one controlled phase to label qubit $j$ (in RR, merged with $\mathbf S_L^\dagger$ and the label Hadamards, a flag-controlled NOT for $s=1$) and uncomputes the flag by measurement (Appendix~\ref{app:prep}). A membership test sets its flag, a reset label qubit, by two multi-controlled NOT gates, $2(d-1)$ ANDs. The parity measurement uses at most $d-1$ CNOT gates and one Hadamard gate (gate sequence in \cite{coderepo}). Each step also takes $O(d)$ single- and two-qubit Clifford gates. The gate $\mathbf T_F$ takes at most $2^b$ multi-controlled $\pi$-phases of $\max\{0,b-2\}$ ANDs each. Table~\ref{tab:gates} multiplies these by the numbers of kicks and tests. RR kicks all $R$ registers with each of its $2Nb$ reports (Theorem~\ref{thm:prep}(iii)). CB applies $2RM_{\rm ser}$ kicks in serial mode and $2RM_{\rm sh}$ in shared mode, both $O(R^2Nb^2/\delta_p)$ because $\chi_{R,N}=\Theta(R)$. Each register in epoch B tests at most the $|M_{c_k}|\le N/2$ edges, or $K$-hop pairs, of its class. Choosing $R=112$, $\delta_p=0.005$ and $b=4$ gives $2R=224$ kicks per raw bit in RR and about $5.2\times10^7$ (serial) or $6.8\times10^7$ (shared) in CB. The direct phase of OC in compatible address order and the HM loading after a record buffer use the unary iteration below, at most $N-1$ ANDs per register. The classical controller uses $O(1)$ operations on $O(\log N)$-bit words per kick, edge and test, and $O(R)$ for the decision. The classical receivers in Table~\ref{tab:compare} likewise use $O(1)$ such word operations per arriving report and per arriving edge.

\begin{table}[!t]
\caption{Gate and classical operation counts for fixed $b$}\label{tab:gates}
\centering\scriptsize\setlength{\tabcolsep}{1pt}
\begin{tabular}{@{}lccc@{}}
\toprule
Receiver & Epoch-A gates & Epoch-B gates & Classical operations\\
\midrule
RR, shared & $O(RNb\log N)$ & $O(RN\log N)$ & $O(RNb+|E|)$\\
CB, serial or shared & $O(R^2Nb^2\log N/\delta_p)$ & $O(RN\log N)$ & $O(R^2Nb^2/\delta_p+|E|)$\\
OC, direct phase & $O(RN)$ & $O(RN\log N)$ & $O(RN+|E|)$\\
Buffer + HM & $O(RN)$ & $O(RN\log N)$ & $O(n_{\rm rep}+RN+|E|)$\\
Store all, site cache & -- & -- & $O(n_{\rm rep}+|E|)$\\
\bottomrule
\end{tabular}
\par\smallskip\parbox{\columnwidth}{\scriptsize Gates: logical ANDs and single- and two-qubit Clifford gates, of the same order; the small CB kick angles add a rotation-synthesis factor; OC assumes compatible address order (otherwise up to $O(RN\log N)$ epoch-A gates). Classical: operations on $O(\log N)$-bit words, a $\mathrm{GF}(2^d)$ product counting as one. $n_{\rm rep}$: total reports, $Nb$ times the reports per raw bit of Table~\ref{tab:compare}.}
\end{table}

\emph{Factory rates.} Let $r_{\rm in}$ be the input rate in raw bits/s, each raw bit counted once however many reports carry it, and $\kappa=A_N/N$ for a schedule with per-register logical-AND work $A_N$ in a window of $Nb$ bits. Each logical AND uses one controlled-controlled-$\mathbf Z$ (CCZ) state from $n_{\rm f}$ factories per register of period $\tau_{\rm f}$ seconds. Average work keeps up if:
\begin{equation}
\tau_{\rm f}\le n_{\rm f}b/(\kappa r_{\rm in})\label{eq:taureq}
\end{equation}
Every schedule that keeps up satisfies \eqref{eq:taureq} for positive work and no initial inventory. This alone does not guarantee keeping up. The direct phase of Remark~\ref{rem:direct} with unary iteration \cite[Sec.~III.A]{Babbush2018} in compatible address order uses at most $N-1$ ANDs per register, so $\kappa\le1-1/N<1$. The kicks of RR use $2b(d-1)$ ANDs per site and register, and those of CB use $d-1$ per report applied to a register. Choosing $n_{\rm f}=1$, $b=4$ and the streams S1--S3 of Section~\ref{sec:massive} ($10^6$, $10^8$ and $10^{10}$ bits/s) gives \eqref{eq:taureq} periods of $4\,\mu$s, $40$\,ns and $400$\,ps. An illustrative surface-code factory takes $5.5\,d_{\rm c}$ code cycles per CCZ state \cite[p.~1]{GidneyFowler2019}. This is about $10^{-4}$\,s with $1\,\mu$s cycles \cite[p.~1]{GidneyEkera2021} and distance $d_{\rm c}\approx20$, or $25$ to $2.5\times10^5$ times these periods. These ratios compare work bounds, not runtimes. Runtimes depend on fidelity, routing, storage and a controller applying report-dependent gates at the stream rate \cite{Markidis2026}.

\section{Simulation settings and standard errors}\label{app:sensor}\label{app:jackknife}
\emph{Graph.} The sensor-network graph of Fig.~\ref{fig:sensor} is a weighted random geometric graph. It has $N=1024$ uniformly drawn points on the unit torus. Points at periodic distance $\varrho\le r_{\rm g}=0.05$ share an edge of weight $w=\exp(-\varrho^2/(2\sigma_w^2))$, $\sigma_w=r_{\rm g}/2$. Its generator seed, colouring and statistics, the sensor map and the settings of Figs.~\ref{fig:sensor}(a) and (b) are documented in the README of the code capsule \cite{coderepo}.

The RMSE standard errors in Figs.~\ref{fig:e2e}(b) and \ref{fig:scaling}(b) are square roots of delete-one jackknife variances \cite[Remark~3.2]{Boucheron2013}. These include the finite-sample prefactor (replicate count minus one over that count). Each statistic omits one replicate and recomputes normalized RMSE at fixed true energy and normalization. No confidence-interval coverage is asserted.

The fixed-realization error bars of Fig.~\ref{fig:e2e}(a) are empirical standard deviations with divisor $40$. Figure~\ref{fig:e2e}(b) has estimated standard errors at most $\figFiveBiasSE$ (signed relative bias before taking its absolute value) and $\figFiveRmseSE$ (relative RMSE). The signed-bias one-standard-error interval crosses zero at $M=5\times10^5$. Figure~\ref{fig:scaling}(a) uses estimated standard errors $s/\sqrt n$. Here $s$ is the sample standard deviation (divisor $n-1$) of the $n$ estimates ($n=100$ at $N\le2^{18}$ and $n=20$ at $N=2^{20}$). Figure~\ref{fig:scaling}(b) uses delete-one standard errors of the normalized RMSE, at most $\figSixRmseSE$. Figure~\ref{fig:sensor}(a) has estimated standard errors at most $\figSensorBiasSE$ for the signed relative bias and $\figSensorRmseSE$ for the relative RMSE (delete-one, as above). Figure~\ref{fig:sensor}(b) has an estimated standard error of the mean Wasserstein errors at most $\figSensorWassSE$ graph-frequency units (sample variances). All uncertainties are conditional on the fixed sensor maps.

\section{Gate-level circuit simulations}\label{app:gatelevel}
\emph{Circuits, simulators and settings.} The circuits, the two simulators of Section~\ref{sec:gatelevel}, their validation and the settings of the runs (maps, classes, block lengths and seeds) are documented in the README of the code capsule \cite{coderepo}.

\emph{Register sweep.} Unit weights give $Y_k=2C$ exactly when $r_k=1$. This event has probability $\cE_G/(2C)$ by Lemma~\ref{lem:outcome}. The number of such scores is binomial for ideal registers. A map\textquotesingle s decision error is its tail beyond $R\tau_{\rm m}/(2C)$, upper for smooth and lower for rough maps. Table~\ref{tab:gatesweep} lists the smallest $R$ with decision errors at most $\delta_s$ for CB and RR registers and this law. The RR error counts agree with this law (smallest two-sided $p$ per configuration $0.015$--$0.14$). The CB error counts agree with this law evaluated at the mean score \eqref{eq:meanrho} of each prepared state ($p\ge0.004$). Thus preparation affects the decisions through its bias. One decision across all trials at or above the counts of \eqref{eq:Rbern} differed from the classical one (a CB false alarm at $N=64$, $R=1024$). Choosing $\kappa_bv=0.05$ ($M/N\approx836$) gives the distance bound $\delta_p=2\chi_{R,N}v$ of Theorem~\ref{thm:prep}(ii). It is $0.06$ at $R=1$. It exceeds one from $R=16$ and is $53$ at $R=819$. Achieving $\delta_p=0.05$ at $R=819$ would take $M/N\approx8.8\times10^5$. Testing at $N=2^{10}$ included $2000$ extra trials after one RR count of the first $1000$ deviated. Its $p=0.0008$ at $R=16$ was the smallest of the $96$ RR $p$-values up to $N=2^{16}$. A union bound gives at most $0.08$; the added trials give $p=0.57$. The RR register RMSE is within $5\%$ of $\sqrt{(\E[Y_k^2]-\cE_G^2)/R}$ at every size and $R$, and that of CB registers at $\kappa_bv=0.01$ within $6\%$.

\begin{table}[!t]
\caption{Smallest register counts with decision error at most $\delta_s$ in the gate-level runs}\label{tab:gatesweep}
\centering\scriptsize\setlength{\tabcolsep}{4pt}
\begin{tabular}{@{}rrrcc@{}}
\toprule
$N$ & $\kappa_bv$ & trials & $\delta_s=0.05$ & $\delta_s=0.01$\\
\midrule
$2^{6}$ & $0.05$ & $1000$ & $256/128/128$ & $512/256/256$\\
$2^{10}$ & $0.05$ & $3000$ & $128/128/128$ & $256/256/256$\\
$2^{12}$ & $0.05$ & $1000$ & $128/128/128$ & $256/256/256$\\
$2^{12}$ & $0.01$ & $1000$ & $128/128/128$ & $256/256/256$\\
$2^{14}$ & $0.05$ & $1000$ & $128/128/128$ & $128/256/256$\\
$2^{16}$ & $0.05$ & $500$ & $64/64/128$ & $256/128/256$\\
$2^{19}$ & $0.05$ & $20$ & $128/64/128$ & $256/64/256$\\
\bottomrule
\end{tabular}
\par\smallskip\parbox{\columnwidth}{\scriptsize Entries: smallest $R$ of $16,32,\ldots,2048$ ($\le1024$ at $N=2^{19}$) with decision error at most $\delta_s$ under both hypotheses, for CB registers / RR registers / the binomial law of ideal registers; \eqref{eq:Rbern} gives $819$ and $1259$. An entry at $N=2^{19}$ means at most one error ($\delta_s=0.05$) or none ($0.01$) in $20$ trials; CB or RR entries below the binomial one are within its sampling error.}
\end{table}

\emph{Memory.} The controller holds $26$--$52$ live classical bits in epoch A and $802$--$828$ in epoch B at $R=256$ ($6{,}184$--$6{,}210$ at $R=2048$). None crosses the cutoff. The simulated receiver's $R(d+b)+1$ qubits fall below the $Nb$ record bits from $N=2^{10}$ at $R=256$ and from $N=2^{14}$ at $R=2048$. A report kick costs $d-1$ logical ANDs and a membership test $2(d-1)$, with a clean ladder of $d-1$ ancillas.

\section*{Acknowledgment}
The numerical calculations were supported in part by the national high-performance computing resources of T\"{U}B\.{I}TAK ULAKB\.{I}M TRUBA. The author conceived the problem, the late-graph setting and the receiver, and directed the analysis, the experiments and the writing. He used Claude (Anthropic) and ChatGPT (OpenAI) as assistants under his direction for four specific tasks: (i)~helping to formulate the formal results and drafting the detailed derivations of their proofs and the calculations in Appendices~A--G; (ii)~writing, to his specifications, the Python simulation and plotting code behind Section~\mbox{\ref{sec:numerics}}, Figs.~\ref{fig:memory}--\ref{fig:gatelevel} and Appendices~H and~I, which he tested and then ran to obtain all reported numerical results; (iii)~producing the TikZ code of Figs.~\ref{fig:overview}--\ref{fig:circuit}; and (iv)~drafting text revisions, cross-checking the manuscript against the code, and editing its language. The derivations were verified step by step by independent scripts, by an independent audit with Gemini (Google), and by the author's own review. The author takes full responsibility for the entire content of this paper.

\section*{Data and Code Availability Statement}
The Python implementation and the simulation data are available on Code Ocean as a compute capsule \mbox{\cite{coderepo}}. The capsule replots the simulated figures of the article from stored data and contains the generators of all simulations.

\end{document}